%% file: fstair.tex
\documentclass[12pt]{article}
\usepackage[totalwidth=460pt,totalheight=610pt]{geometry}
\usepackage{amsmath,amssymb,epsf,graphics,graphicx,mathrsfs}
\usepackage{enumerate}
\usepackage{color}
\usepackage{psfrag}
\newcommand{\CM}{{\cal M}}

\newcommand{\RR}{\mathbb{R}}

\newcommand\blank[1]{}

\newcommand{\fract}[2]{{\textstyle\frac{#1}{#2}}}

\newcommand\eq{\begin{equation}}
\newcommand\en{\end{equation}}
\newcommand\bea{\begin{eqnarray}}
\newcommand\eea{\end{eqnarray}}
\newcommand\nn{\nonumber}
\newcommand\ba{\(\begin{array}}
\newcommand\ea{\end{array}\)}
\newcommand{\resection}[1]{\setcounter{equation}{0}\section{#1}}

\newcommand\ep{\varepsilon}

\begin{document}
\begin{titlepage}
\vskip 0.5cm
\begin{flushright}
DCPT--14/81  \\
December 2014
\end{flushright}
\vskip 1.2cm
\begin{center}
{\Large{\bf Form factor relocalisation and interpolating\\[5pt]
renormalisation group flows from the staircase model}}
\end{center}
\vskip 0.8cm \centerline{Patrick Dorey$^{1}$, Guy Siviour$^{1}$ and
G\'abor Tak\'acs$^{2,3}$} \vskip 0.9cm

\vskip 0.3cm \centerline{${}^{1}$\sl\small Department of Mathematical
Sciences, Durham University,} \centerline{\sl\small South Road, Durham
DH1 3LE, United Kingdom}

\vskip 0.3cm \centerline{${}^2$\sl\small 
MTA-BME “Momentum” Statistical Field Theory Research Group,}
\centerline{\sl\small 1111 Budapest, Budafoki út 8, Hungary}
\vskip 0.3cm \centerline{${}^{3}$\sl\small 
Department of Theoretical Physics, }
\centerline{\sl\small 
Budapest University of Technology and Economics,}
\centerline{\sl\small 
1111 Budapest, Budafoki út 8, Hungary
}
\vskip 1cm 
\centerline{E-mails:}
\centerline{p.e.dorey@durham.ac.uk,  gbsiviour@gmail.com, 
takacsg@eik.bme.hu
}

\vskip 1.25cm
\begin{abstract}
\noindent
We investigate the staircase model, introduced by
Aliosha Zamolodchikov through an
analytic continuation of the sinh-Gordon S-matrix to describe
interpolating flows between minimal models of conformal field theory
in two dimensions.
Applying the form factor expansion and the $c$-theorem, we show that 
the resulting $c$-function has the same physical content as that
found by Zamolodchikov
from the thermodynamic Bethe Ansatz. This turns out to be a
consequence of a nontrivial underlying mechanism, which leads to an
interesting localisation pattern for the spectral integrals giving the
multi-particle contributions. We demonstrate several aspects of this
form factor relocalisation, which suggests a novel approach to the
construction of form factors and spectral sums in integrable
renormalisation group flows with non-diagonal scattering.
\end{abstract}

%

\end{titlepage}
\setcounter{footnote}{0}
\def\thefootnote{\fnsymbol{footnote}}
%
\resection{Introduction: the staircase models}
Of all integrable two-dimensional quantum
field theories admitting a Lagrangian description,
the sinh-Gordon model is the simplest to define.
Nevertheless its properties
continue to surprise, and it
is far from being completely understood. 
In 1991, in a paper that circulated for many years 
in preprint form before it was
finally published in 2006 \cite{Zamolodchikov:1991pc},
Aliosha Zamolodchikov pointed out a further curious feature: 
the S-matrix of the model
admits an interesting continuation from its self-dual point
to certain complex values
of the coupling. The S-matrix remains real-analytic,
being real for purely-imaginary values of the rapidity, but acquires a
couple of `resonance poles' located just off the physical sheet.
As a scattering theory this continued S-matrix appears to make sense, 
even though the Lagrangian description of the resulting theory is
somewhat obscure.
Zamolodchikov chose instead
to study the properties of the model via the
thermodynamic Bethe ansatz (TBA) method, which gives access to the 
finite-volume ground-state
energy of a quantum field theory taking as
its only input the exact S-matrix. This revealed an
intriguing structure: as a parameter $\vartheta_0$ encoding the
`distance' of the continued S-matrix from the self-dual point was taken to
infinity, the ground-state energy found by plugging the continued
S-matrix into the TBA exhibited a sequence of scaling
behaviours approximating with increasing accuracy those of the minimal
conformal field theories $\CM_p$, one after the other. The crossovers
between these regions approximated, again with increasing accuracy as
$\vartheta_0$ increased, the $\CM_p\to\CM_{p-1}$
interpolating flows that had previously been found perturbatively in
\cite{Zamolodchikov:1987ti,Ludwig:1987gs}
and analysed exactly using the TBA in
\cite{Zamolodchikov:1991vx,Zamolodchikov:1991vh}. 

Zamolodchikov interpreted these results as suggesting the existence of
a family of integrable quantum field theories with
what he dubbed `roaming' renormalisation group (RG) trajectories, passing
close by each of the minimal models in turn before finally flowing to
massive theories in the far infrared. The general idea is illustrated
in figure \ref{stairflow} below; each black dot represents an RG fixed
point described by a conformal field theory, and the larger the 
parameter $\vartheta_0$ becomes, the nearer the
trajectory passes to each fixed point, and the longer it spends there.
In the immediate neighbourhood of each fixed point, the model can be
described by a combination of a relevant ($\phi_{13}$) and an
irrelevant ($\phi_{31}$) perturbations of the corresponding conformal
field theory, a picture which was subsequently verified perturbatively
to be consistent with the multiply-hopping flows by L\"assig 
\cite{Lassig:1991ab}. 

\begin{figure}[h]
\[
\begin{array}{c}
\resizebox{10cm}{!}{
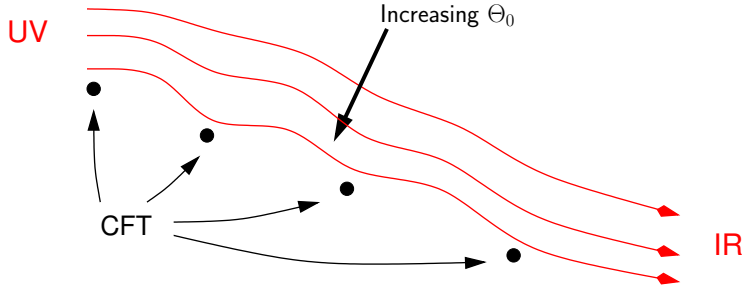
}
\end{array}
\]
\caption{\small A collection of roaming renormalisation group
flows, parameterised by $\vartheta_0$}
\label{stairflow}
\end{figure}

Zamolodchikov's paper led to a series of generalisations 
\cite{Martins:1992ht,Dorey:1992bq,Martins:1992sx,Dorey:1992pj}, with
even more elaborate one-parameter
families of RG flows emerging from 
proposed sets of TBA equations, each visiting
infinitely-many fixed points in a suitable limit.
Similar behaviours were then predicted for the 
Homogeneous Sine-Gordon (HSG) models 
\cite{Miramontes:1999hx} and then confirmed by
a TBA analysis 
\cite{CastroAlvaredo:1999em,Dorey:2004qc}, though in these cases the
number of fixed points visited is always finite.
The pattern in all cases is of successive hops between a sequence of
conformal
field theories with decreasing central charges, punctuated by ever-longer
`pauses', or plateaux, near to each conformal field theory.
For this reason the roaming theories are often called staircase
models.

Given the ubiquity of the roaming phenomenon, it seems worthwhile to
understand it in more depth, in particular to gather more evidence as
to whether the roaming trajectories encode the properties of
bona fide local quantum field theories, or are merely 
mathematical artefacts of the conjectured TBA equations. Further
motivation comes from the fact that the S-matrices associated with the
roaming trajectories are often far simpler than those of
the interpolating flows that they ultimately come to
approximate: for example, Zamolodchikov's
original staircase S-matrix is
diagonal, while the massless S-matrices
associated with the $\CM_p\to\CM_{p-1}$
interpolating flows \cite{Fendley:1993xa}
are non-diagonal and significantly more
complicated. This has recently been used to conjecture exact
equations describing combined bulk and boundary flows between minimal
models \cite{Dorey:2010ub}, confirming and extending previous
perturbative results 
\cite{Fredenhagen:2009tn}.

In this paper we consider Zamolodchikov's staircase model not through the
TBA approach, but rather via the form factors that follow from its
S-matrix. These give access to (Sasha) Zamolodchikov's $c$-function 
\cite{Zamolodchikov:1986gt}
for the model, which we find through both
numerical and analytical treatments behaves exactly 
as expected in the 
$\vartheta_0\to\infty$
limit. 
For the HSG models, the form factor expansion of
Zamolodchikov's $c$-function was studied numerically in 
\cite{CastroAlvaredo:2000ag,CastroAlvaredo:2000nr} (see also 
section 7 of \cite{Dorey:2004qc}), and the evidence found therein
for a roaming behaviour was an important motivation for our work.
However these earlier papers did not provide an analytic explanation for
the observed phenomena, something which is of independent interest.
In particular, the staircase limit allows us to extract a
set of `effective' form factors for the interpolating flows. For the
tricritical Ising to Ising flow these reproduce the diagonal
massless form factors found
by Defino, Mussardo and Simonetti
\cite{Delfino:1994ea}, while for flows further up the staircase an
interesting picture incorporating extra `magnonic'
contributions emerges, hinting at further universal structures in the
general form-factor description of perturbed conformal field theories.

In the next section recall some further details of Zamolodchikov's
staircase model and its treatment via the TBA. In section 3 we
review the form factor approach and present some numerical
results concerning the form factor treatment of the staircase limit.
These results are supported by analytic arguments in section 4, while
section 5 concludes the paper with some speculations and suggestions
for further work.

\resection{A short review of the staircase TBA system}\label{sec:stairTBA}
%
%

\begin{figure}[h]
\[
\begin{array}{c}
\includegraphics[width=180pt]{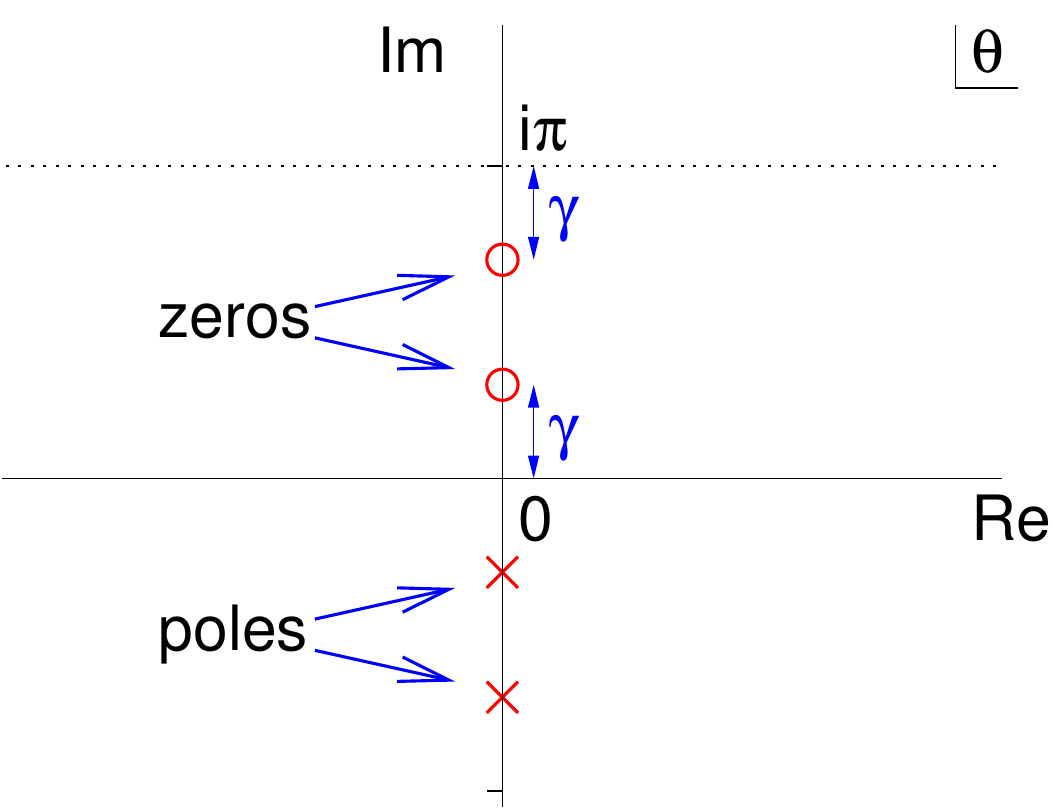}
\end{array}
\]
\caption{\small
The analytic structure of the sinh-Gordon S-matrix: there are
two zeros on the physical strip $0\le\Im m\,\theta\le \pi$, and two
poles on the unphysical strip $-\pi\le\Im m\,\theta\le 0$.}
\label{shGS}
\end{figure}

The sinh-Gordon model is a theory of a single scalar field $\Phi$,
with a classical action depending on a mass scale $M$ and coupling $b$:
\begin{equation}
\mathcal{A}=\int
d^{2}x\left(\frac{1}{2}\partial_{\mu}\Phi\partial^{\mu}\Phi-\frac{M^{2}}{b^{2}}
\cosh b\Phi\right)\; .
\label{eq:sinhGaction}
\end{equation}
The model is integrable, with one of the simplest non-trivial exact S-matrices
known: in terms of the parameter
$\gamma:=\pi b^2/(8\pi+b^2)$, it is
\eq
S(\theta)=
\frac{\sinh\theta-i\sin\gamma}{\sinh\theta+i\sin\gamma}\; .
\label{eq:sinhGSmatrix}
\en
This function has a pair of zeros at $i\gamma$ and
$i(\pi{-}\gamma)$
in the physical strip of the complex rapidity plane, and no
physical-strip poles;
as $b$ varies from $0$ to $\infty$, $\gamma$ moves from $0$ to
$\pi$ and the zeros swap over, reflecting the strong-weak coupling
duality of the model under $b\to 1/b$, $\gamma\to \pi-\gamma$.
The situation is depicted in figure \ref{shGS}.

\begin{figure}[h]
\[
\begin{array}{c}
\includegraphics[width=180pt]{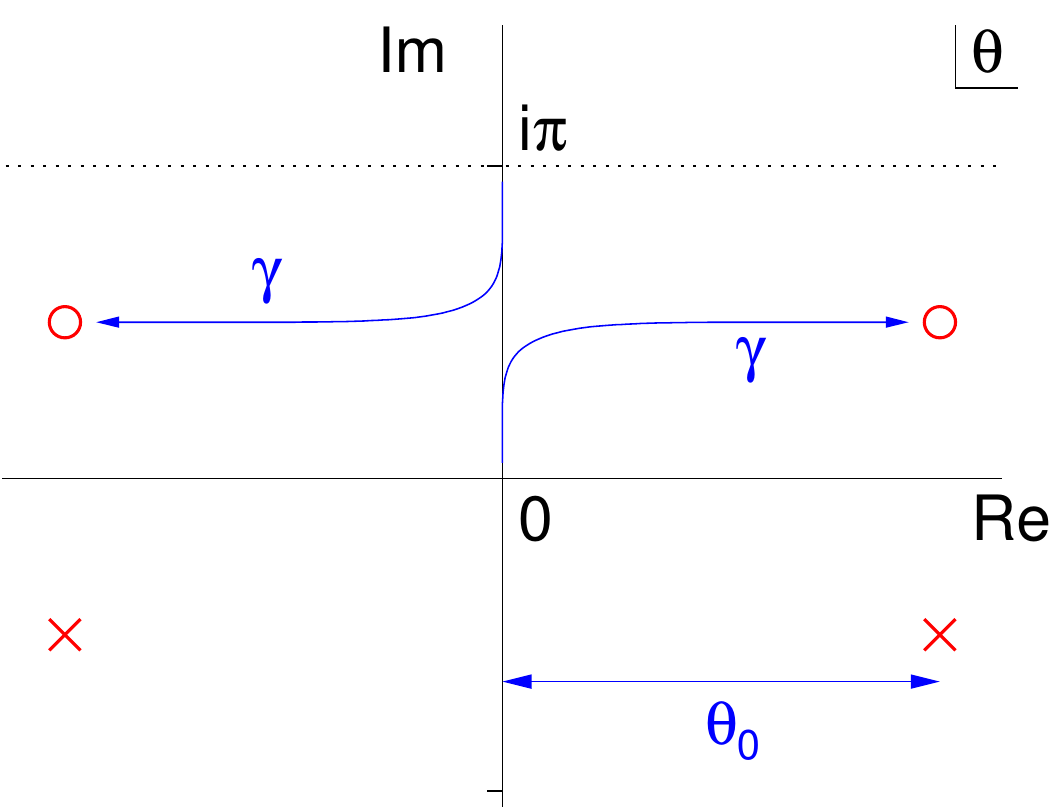}
\end{array}
\]
\caption{\small
The S-matrix continuation from sinh-Gordon to staircase: the
zeros and poles move a distance $\pm\vartheta_0$ parallel to the real
axis from their positions at the self-dual point.}
\label{shGScont}
\end{figure}

The roaming S-matrix is obtained from (\ref{eq:sinhGSmatrix}) 
by analytically continuing away from
the self-dual point $\gamma=\pi/2$, setting
\begin{equation}
\gamma=\frac{\pi}{2}\pm i\vartheta_0
\label{eq:sinpibper2roaming}
\end{equation}
with $\vartheta_0$ real, and
letting $\vartheta_0$ tend to infinity in the staircase limit. The
resulting S-matrix is depicted in figure \ref{shGScont}. It
is still real-analytic, but now has a pair of forward and crossed
channel `resonance
poles' on the unphysical sheet, with real parts $\pm\vartheta_0$.

Whether this continuation  makes sense at the level of the
action (\ref{eq:sinhGaction}) is an open question (though see some
speculations in the final section of \cite{Dorey:1992bq}). However
as an abstract S-matrix encoding the scattering of
asymptotic multiparticle
states there are no immediate problems, and under the assumption
that this is the S-matrix of underlying  massive field theory 
one can try to
discover further properties of this theory using standard
methods of integrable models, in particular the TBA and form-factor
approaches.

The TBA gives the ground-state energy $E_0(R)$
of an integrable QFT on a circle of circumference $R$ in terms of the
solutions of one or more  non-linear integral (TBA) equations. For a theory
with a single massive particle of mass $m$, such as the sinh-Gordon or
staircase model, there is just one TBA equation, for a function
$\ep(\theta)$ known as the pseudoenergy:
\eq
\ep(\theta)=r\cosh(\theta)-
\int_{\RR}\phi(\theta-\theta')L(\theta')\,d\theta'
\label{eq:TBA}
\en
where $L(\theta)=\log(1+e^{-\ep(\theta)})$,
$\phi(\theta)=-\frac{i}{2\pi}\frac{d}{d\theta}\log S(\theta)$, and
$r=MR$ is the dimensionless system size in units of $1/M$, the correlation
length. The ground-state energy is
then expressed in terms of a scaling function $c_{\rm
eff}(r)$ known as the effective central charge,
\eq
E_0(R)=-\frac{\pi}{6R}\,c_{\rm eff}(r)\,,
\label{eq:Eceff}
\en
as
\eq
c_{\rm eff}(r)=\frac{3}{\pi^2}\int_{\RR}r\cosh(\theta)
L(\theta)\,d\theta\,.
\label{eq:ceff}
\en
For a unitary conformal field theory with central charge $c$, the 
scaling of the 
ground-state energy is
such that $c_{\rm eff}(r)$ is a constant, equal to $c$.
On the other hand, if a theory not
conformal but at some
length-scale $r$ is close to a conformal field theory, then 
$c_{\rm eff}(r)$ near that same value of $r$
should be close to the corresponding central
charge. For example, as $r\to 0$ the effective central charge of a
conformal field theory perturbed by some relevant operator should tend
to the central charge of the unperturbed model; but more generally,
any pause in the evolution of $c_{\rm eff}(r)$ at an intermediate scale
is evidence that the renormalisation group trajectory of the model
is passing close by a CFT, with central charge equal to the
approximately-constant value of
$c_{\rm eff}(r)$ at that scale.

\begin{figure}[h]
\[
\begin{array}{cc}
\vartheta_0=0: &\quad
\parbox[c]{180pt}{\includegraphics[height=60pt]{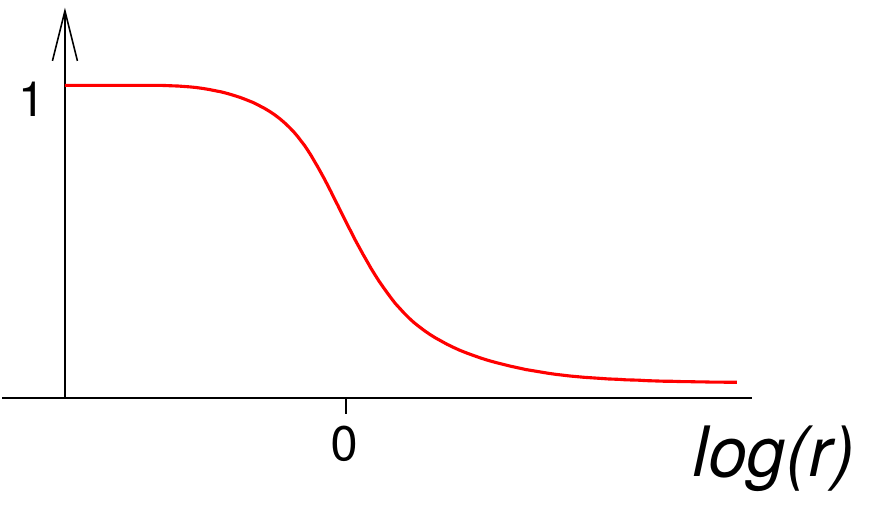}}
\\
\vartheta_0=10: &\quad
\parbox[c]{180pt}{\includegraphics[height=60pt]{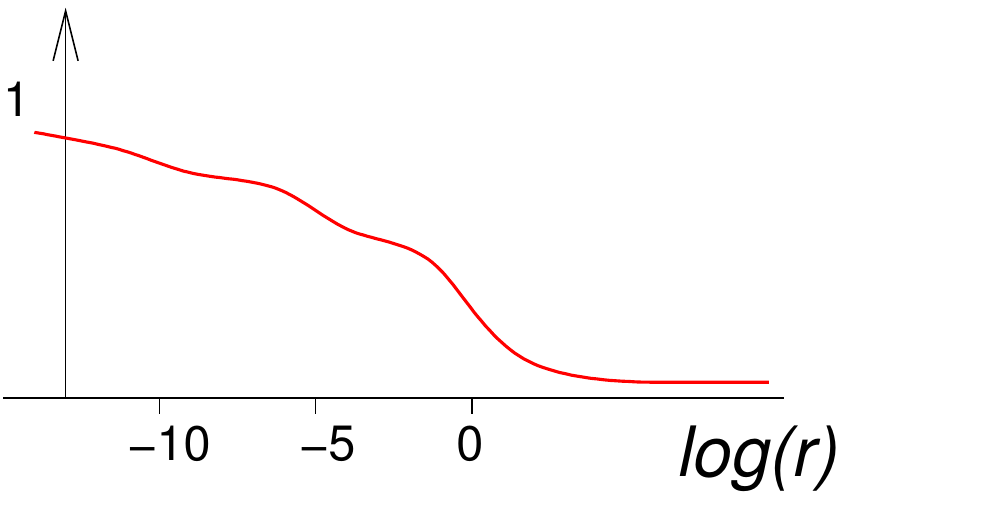}}
\\[3pt]
\vartheta_0=20: &\quad
\parbox[c]{180pt}{\includegraphics[height=60pt]{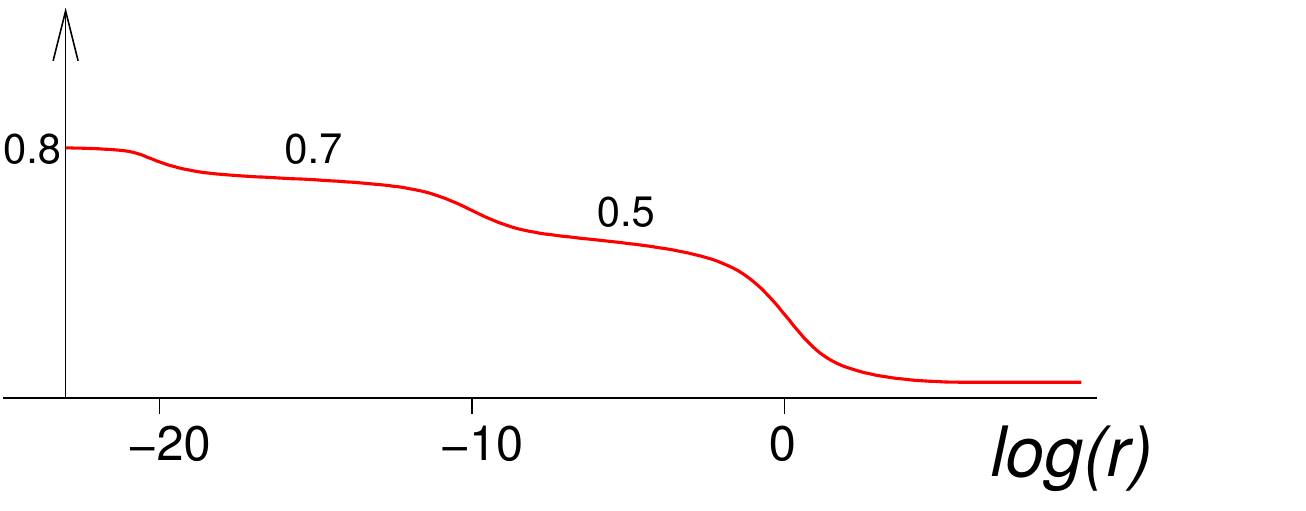}}
\end{array}
\]
\caption{\small
The effective central charge of the
staircase model for various values of $\vartheta_0$.}
\label{fig:stairplots}
\end{figure}

For the staircase model, this is exactly what Zamolodchikov observed.
Plots of $c_{\rm eff}$ as a function of $\log(r)$ show the same
UV and IR limits as those for the sinh-Gordon model, namely
$1$, the central charge of a
single free boson, and $0$, the infrared value always found in a theory
with no massless degrees of freedom at long distances. But as
$\vartheta_0$ grows they
acquire an increasingly-pronounced series of plateaux at intermediate
scales, of widths $\vartheta_0/2$.  The heights of
these plateaux are the central charges of the
unitary $c<1$ minimal models, as illustrated in figure
\ref{fig:stairplots}. These plots imply precisely the RG flows
sketched in
figure~\ref{stairflow} above, with the increasing
length of RG time spent on each plateau 
indicating that the corresponding RG trajectories get nearer and
nearer to the RG fixed points as $\vartheta_0$ increases.

\begin{figure}[h]
\[
\begin{array}{c}
\includegraphics[width=180pt]{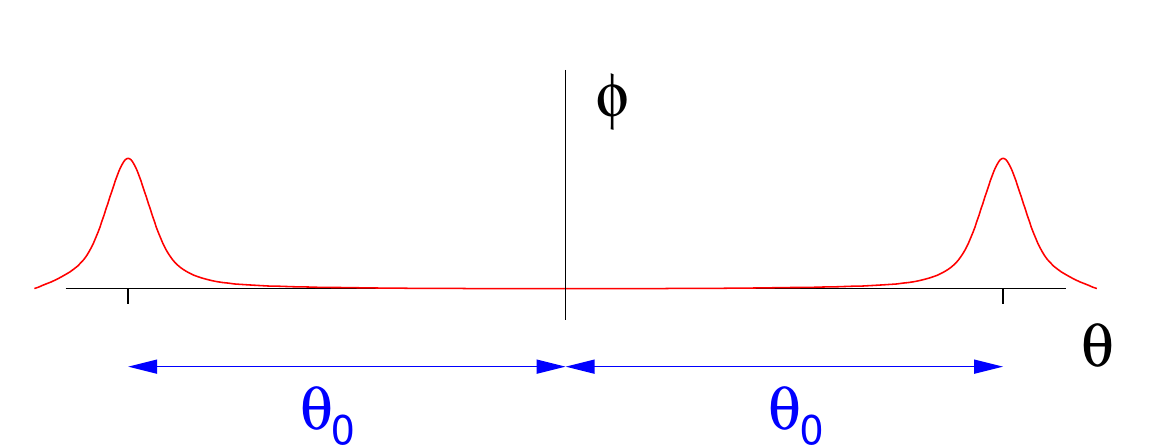}
\end{array}
\]
\caption{\small
The TBA kernel function $\phi(\theta)$ for the staircase
model.}
\label{fig:phi}
\end{figure}

To understand how this pattern emerges from the
TBA equation (\ref{eq:TBA}), consider the form of the
`kernel function' $\phi(\theta)$, which in terms of the parameter
$\vartheta_0$
is
\eq
\phi(\theta)=
\frac{1}{2\pi\cosh(\theta+\vartheta_0)}
+\frac{1}{2\pi\cosh(\theta-\vartheta_0)}~.
\en
When $\vartheta_0$ is large this function `relocalises' into a pair of 
disconnected
peaks of width of order 1, situated at $\theta=\pm\vartheta_0$ on the
real axis, as shown 
in figure \ref{fig:phi}.

In turn, the relocalised kernel means that the integral in
the TBA equation
(\ref{eq:TBA}) couples $\ep(\theta)$ not to $\ep(\theta')$ with
$\theta'\approx\theta$, but rather to its values at
$\theta'\approx\theta\pm\vartheta_0$. However one must also
consider the effect of the
driving term $r\cosh(\theta)$ in the TBA equation: for
$|\theta|\gg\log(1/r)$ this dominates, causing $\ep(\theta)$ to be large
and $L(\theta)$ small. Nontrivial behaviour of $\ep(\theta)$
therefore only occurs in the region 
$-\log(1/r)\lessapprox\theta\lessapprox\log(1/r)$, and
this behaviour depends crucially on how many times the shift
$\vartheta_0$ fits into this interval. This number changes whenever
$2\log(1/r)\approx (k{-}1)\vartheta_0$, or $\log(r)\approx
-(k{-}1)\vartheta_0/2$, 
explaining the
additional steps in the staircase seen in figure
\ref{fig:stairplots}.\footnote{A remark on notation: most of our focus
here and below is on behaviours as $\vartheta_0\to\infty$.
Accordingly, $\theta'\approx\theta$ means that $\theta'-\theta$
remains finite (ie of order 1) as $\vartheta_0\to\infty$, while
$\theta'\gg\theta$ means that $\theta'-\theta$ grows to infinity in
this same limit.
}

Note that to see the limiting behaviour of $c_{\rm eff}(r)$ on any 
individual step, it is not enough simply to take $\vartheta_0$ to
infinity: $r$ must be rescaled to zero simultaneously, otherwise the
step will be missed. More precisely, to see the $k^{\rm th}$ step, set
\eq
r=\tilde r\,e^{-(k{-}1)\vartheta_0/2}
\en
and then let $\vartheta_0$ tend to infinity with $\tilde r$ remaining
finite. In this limit the TBA equation couples the neighbourhood
$\theta\approx\log(1/r)\approx (k{-}1)\vartheta_0/2$ to $\theta\approx
(k{-}3)\vartheta_0/2$, which is in turn coupled to 
$\theta\approx (k{-}1)\vartheta_0/2$ and to 
$\theta\approx (k{-}5)\vartheta_0/2$ and so on.
As $\vartheta_0$ grows
these neighbourhoods separate, and 
$\ep(\theta)$ is best described by introducing a sequence of
`effective' pseudoenergies, one for each neighbourhood. As explained
in detail in \cite{Dorey:1992bq}, 
as $\vartheta_0\to\infty$ 
the equations governing these
effective pseudoenergies and also the resulting function $c_{\rm
eff}(\tilde r)$ become
exactly the massless TBA equations introduced by Zamolodchikov in 
\cite{Zamolodchikov:1991vx,Zamolodchikov:1991vh}, confirming that, at
least as far as the evolution of the effective central charge is
concerned, the limiting flow of the staircase model is indeed the 
sequence of $\CM_p\to\CM_{p-1}$ flows, with step $k$ corresponding 
to $p=k+2$.

In the following we will show that a similar story holds for the
form factors. The details are a little more intricate, but the
basic idea of relocalisation of integrals, coupled with a form of
double-scaling limit involving $r$ and $\vartheta_0$ to expose the
individual steps, is the same.

\resection{Some form factor phenomenology}
\subsection{Form factors, Zamolodchikov's $c$-theorem and Cardy's sum
rule}

An alternative scale-dependent observable to the finite-size
ground-state energy is the correlation function of two operators
separated by a distance $R$. In a conformal field theory this will
scale as a power of $R$; otherwise if in some regime the theory is
close to a CFT, then approximate power-law scaling should be observed.
However, rather than look for this behaviour directly, Sasha
Zamolodchikov pointed out that 
the correlation functions of the components of the energy-momentum 
tensor can be used to construct a quantity -- 
the Zamolodchikov $c$-function $c(R)$ \cite{Zamolodchikov:1986gt} --
which is constant for a CFT, equal to the theory's central
charge, and which for unitary non-conformal field theories is a
decreasing function of $R$. More precisely, we have
\begin{equation}
\frac{dc(R)}{dR}=-\fract{3}{2}R^3\langle\Theta(R)\Theta(0)\rangle\;.
\label{eq:cflow_equation}\end{equation}
where $\Theta$ is the trace of the energy-momentum tensor. As stressed
by Cardy, it is useful to integrate this last formula to yield the sum rule
\cite{Cardy:1988tj}:
\eq
c(R)= c(\infty)
+\fract{3}{2}\int_{R}^{\infty}\!dR'\,
(R')^3\langle\Theta(R')\Theta(0)\rangle\,.
\label{sumrule1}
\en
In general $c(R)$ is not the same as the effective central charge
discussed in the last section, but
the two functions agree
for critical models, and are equally effective as tools to
analyse an RG flow. 

Form factors allow general correlation functions, and in particular
those appearing in (\ref{sumrule1}), to be expressed
as an infinite sum of multiple integrals; and for an integrable QFT the
integrands can be determined exactly, at least in principle, once the
S-matrix is known.
In a theory with a single massive particle of mass $m$ and
$n$-particle asymptotic states 
$|\theta_1,\dots\theta_n\rangle$
with rapidities $\theta_1$\dots
$\theta_n$,
the form factors of the trace of the energy-momentum tensor $\Theta(x)$ 
are defined as
\begin{equation}
F^{\Theta}_n=\langle 0 |\Theta(0)|\theta_1,\dots\theta_n\rangle\,.
\end{equation}
The two-point function of $\Theta$ can be expressed in terms of
these form factors by inserting a
complete set of states:
\begin{equation}
\langle\Theta(R)\Theta(0)\rangle=
\sum_{n=0}^{\infty}\int_{\RR^n}\!\frac{d\theta_1\dots
d\theta_n}{(2\pi)^nn!}
|F_n^{\Theta}(\theta_1\dots\theta_n)|^2\,e^{-mR\sum_{i=1}^n\cosh\theta_i}.
\label{eq:Theta2ptFF}\end{equation}
Substituting this expansion 
into (\ref{sumrule1}) 
and using the fact that for a massive field theory $c(\infty)=0$,
the value of the $c$-function  is
\eq
c(r)
=3
\sum_{n=0}^{\infty}\int_{\RR^{n}}\!\frac{d\theta_1\dots d\theta_{n}}
{(2\pi)^n(n!)}\,
\frac{6+6rE+3r^2E^2+r^3E^3}{2E^4}\,
|m^{-2}F_{n}^{\Theta}(\theta_1\dots\theta_{n})|^2\,e^{-rE}\; ,
\label{c-sumrule}
\en
where $r=mR$ now denotes the separation of the fields in the two-point
functions defining $c(r)$ in units of the correlation length,
and the dimensionless `energy' $E$ is
\begin{equation}
E=\sum_{i=1}^{n}\cosh\theta_i \; .
\end{equation}
From now on we use units in which $m=1$, and focus on the 
sinh-Gordon and staircase models. For ease of reference, some
key formulae for the sinh-Gordon form factors together with their most
important properties are provided in
appendix 
\ref{app:shGFF}.

\subsection{Brute force numerics}

Under the roaming continuation
(\ref{eq:sinpibper2roaming}),
$\sin\gamma$ is replaced by $\cosh\vartheta_0$
and $\vartheta_0$ is then taken to infinity. Continuing the exact form
factors and
evaluating the form factor sum rule (\ref{c-sumrule}) numerically 
for large values $\vartheta_0$, we can check whether the roaming
behaviours of the TBA results are also seen in correlation functions.


\subsubsection{The Ising flow}
If $r$ is held fixed and nonzero as the limit
$\vartheta_0\rightarrow\infty$ is taken, then all rapidity integrals
in (\ref{c-sumrule}) are effectively confined to a finite (and
$\vartheta_0$-independent) region by the factor $e^{-rE}$.
In such a region the form factors 
(\ref{eq:Theta0exp1}) of $\Theta$ have the limits
\begin{eqnarray}
& & F_2^{\Theta}(\theta_1,\theta_2)=-2 \pi i \sinh\frac{\theta_1-\theta_2}{2} 
\nonumber \\
& & F_{n}^{\Theta}(\theta_1,\dots,\theta_n)=0\quad\mathrm{for}\quad n \neq 2 
\; ,
\label{ffisinglim}
\end{eqnarray}
which coincide with the form factors for the Ising model a.k.a.\ the
free massive 
Majorana fermion. From (\ref{c-sumrule})
one obtains
\cite{Ahn:1993dm}
\begin{equation}
\lim_{r\to 0}\left[\lim_{\vartheta_0\to\infty}c(r,\vartheta_0)\,\right]
=\frac{1}{2}\; ,
\end{equation}
which is the correct value for an Ising fixed point in the ultraviolet.

\subsubsection{Flow from tricritical to critical Ising}

\begin{figure}[ht]
\begin{center}
\includegraphics[scale=0.7]{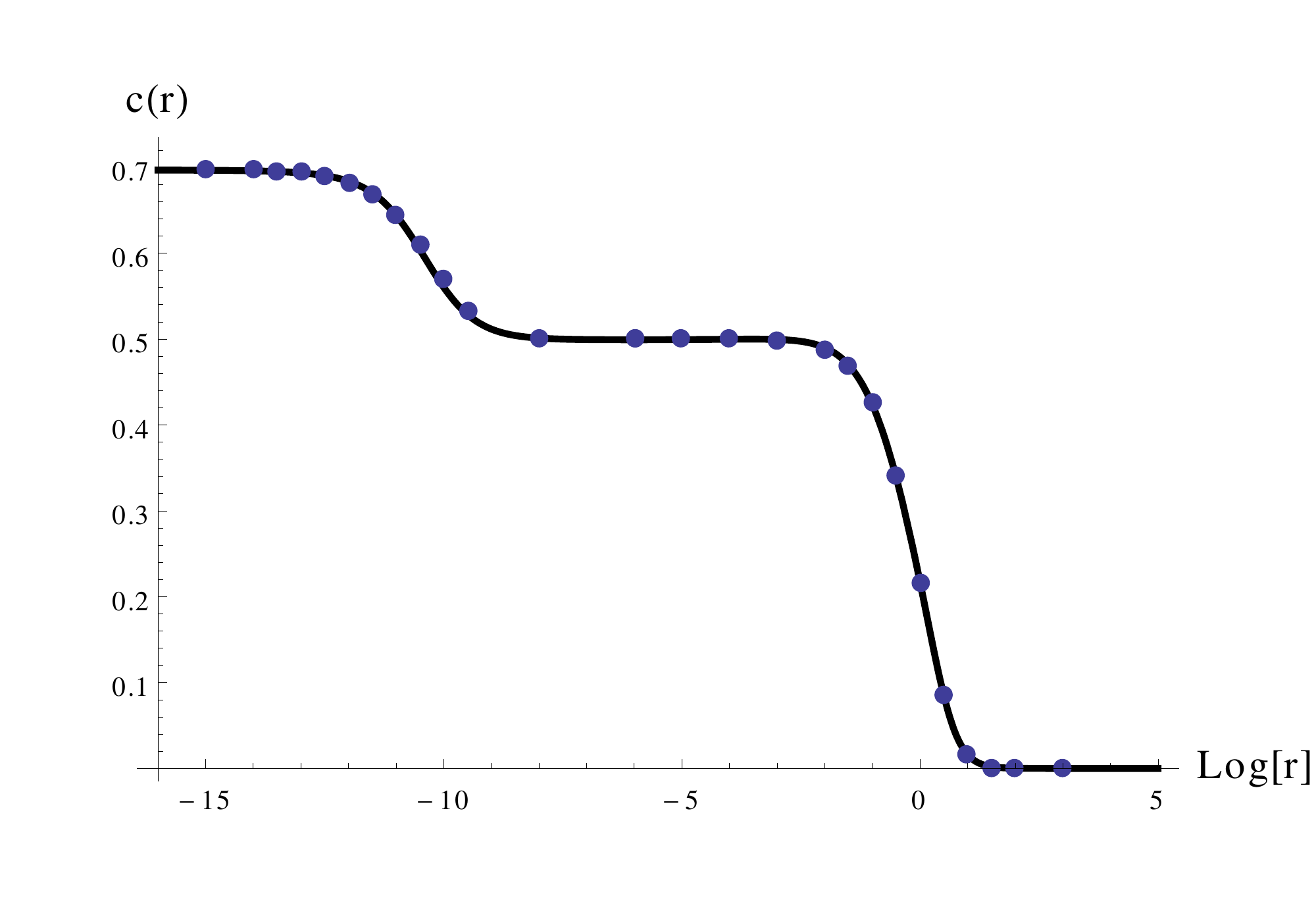} 
\end{center}
\caption{\small
The first two steps of the staircase for 
$\vartheta_0=20$, including 2 and 4 particle contributions. The dots show the 
result of numerical integration, while the curve is a fit 
by an appropriate rational function of $r$.}
\label{imandtimfig}
\end{figure}

To obtain the next step in the staircase, a simple
$\vartheta_0\rightarrow\infty$ limit is not sufficient. Instead, one
must retain the full expressions for the form factors and evaluate the
sum rule for a large but fixed value of $\vartheta_0$, varying the
parameter $r$. The result is shown in figure \ref{imandtimfig}: to
obtain the second step we added the 4-particle contributions which
contribute a central charge difference $\Delta c=0.19767\dots$ (the
result is terminated at the last accurately known digit).  It is
expected that the second plateau must be at the tricritical Ising
value $c=7/10$ so the exact value must be $\Delta c=0.2$; as
demonstrated later, the rest of the central charge difference is
accounted for by terms with more than 4 particles.

Note that the first step from the Ising model with $c=1/2$ appears at 
$\log r\approx 0$, 
while the second one is 
centred at $\log r=\vartheta_0/2$, 
which explains why it cannot be seen in a naive  
$\vartheta_0\rightarrow\infty$ limit. This is a general pattern: 
the $k^\mathrm{th}$ step, 
approximating the flow ${\mathcal M}_{p}\rightarrow{\mathcal 
M}_{p-1}$ with $p=k+2$ when $\vartheta_0$ is large, is centred at
\begin{equation}
 \log r=(k-1)\frac{\vartheta_0}{2}\; .
\end{equation}
This behaviour is in full agreement with that of the TBA system discussed in Section \ref{sec:stairTBA}.

However, higher steps are much harder to calculate to a good accuracy as the 
integrals are difficult even for sophisticated 
multi-dimensional integration algorithms\footnote{To evaluate the rapidity 
integrals, we used the Divonne routine of the Cuba library from Feynarts 
(http://www.feynarts.de/cuba), which implements an adaptive pseudo-Monte Carlo 
algorithm.}. The underlying reasons are the growing dimensionality of the 
integrals, the growing ranges of rapidities which must be considered
when evaluating these integrals,
and
the progressively more complex shapes of the regions within these
ranges where the integrand is significantly different from zero.
To improve this situation, it is necessary to understand the nature of the 
regions to which the integral eventually localises for large 
values of $\vartheta_0$.

\subsection{Cells and localisation}\label{cellsandloc}

The simplest strategy is to divide up the integration space into cells with a 
hypercubic shape of 
size $\vartheta_0/2$, with centres on a hypercubic lattice with the same 
spacing. This choice is suggested by the fact
that the $e^{-rE}$ factor in the integrand
restricts
all rapidity integrals to the ranges
$-\log(1/r)\lessapprox\theta_i\lessapprox\log(1/r)$, while the steps
are expected to occur each time
$\log(1/r)$ passes an integer multiple of $\vartheta_0/2$. An analytic 
justification will be given in 
section \ref{analytic}. We introduce the following notation for the 
hypercubes in rapidity space:
\begin{equation}
[a_1,\dots,a_n]_\alpha:=
\left\{ (\theta_1,\dots,\theta_n)\in\mathbb{R}^n\,:  
a_k \frac{\vartheta_0}{2} - \frac{\alpha}{2}\leq \theta_k
\leq a_k  \frac{\vartheta_0}{2} + \frac{\alpha}{2}\; ,\; k=1,\dots ,n 
\right\}\; ,
\end{equation}
where the integers $a_k$ describe the centre of the cell, while $\alpha$ allows 
for varying its size, 
with $\alpha=\vartheta_0/2$ corresponding to a full elementary cell of the 
lattice.

It is easy to see that cells 
differing by symmetry transformations generated by permutations of the
centre 
coordinates $a_1,\dots,a_n$,
and by changing their signs simultaneously contribute the same amount, therefore 
it is enough to evaluate a 
representative for each sets of equivalent cells. A cell is in normal form if 
$a_1<0$ and the sequence 
$a_1,\dots,a_n$ is monotonically increasing; each equivalence class of cells can 
be obtained by applying 
the symmetries to a cell of normal form\footnote{The reflection composed of 
inverting the signs, and the order of integers takes
a normal cell into another normal cell. As a result, some classes contain one 
cell of normal form which is reflection invariant, 
while other classes contain two cells of normal form which are reflections of 
each other.}.

Performing numerics with the above settings, the following key observations can 
be made:

\begin{enumerate}
 \item While increasing $\vartheta_0$ gives more accurate values for the steps, 
especially for higher ones, for each cell 
 $[a_1,\dots,a_n]_\alpha$ its contribution is essentially saturated by $\alpha$ 
of order 1, which stays the same  as $\vartheta_0$ increases. 
Therefore the index $\alpha$ is omitted from now on, and the contribution of 
cell $[a_1,\dots,a_n]$ is always understood to be the saturated value.
 \item The contribution to the $k^{\rm th}$
 step starts at particle number $n=2k$. It 
is also interesting to note that the bulk of the contribution 
 is already obtained from this level.
 \item Especially for higher steps, the number of potential cells is
very large, 
but only a few of them contribute significantly. For the first four steps 
 and up to 8 particles we list the contributing cells in Table \ref{boxes}. 
The contributions of 'significant' cells tends to a constant, while the 
contributions of others decrease to $0$ in the large  $\vartheta_0$ limit.
Numerically it can be inferred that the asymptotic values of the cell 
contributions (whether zero or finite) are approached exponentially fast in 
$\vartheta_0$.
\end{enumerate}

\begin{table}[h]
\begin{center}
\begin{tabular}{|l||l|l|l|l|}
\hline
Step & 2-particle & 4-particle    & 6-particle          & 8-particle             
    \\
\hline
1    & $[0,0]$    & --            & --                  & --                     
    \\
\hline
2    & --         & $[-1,-1,1,1]$ & $[-1,-1,-1,-1,1,1]$ & $[-1,-1,1,1,1,1,1,1]$  
    \\
     &            &               &                     & 
$[-1,-1,-1,-1,1,1,1,1]$    \\
\hline
3    & --         & --            & $[-2,-2,0,0,2,2]$   & $[-2,-2,0,0,0,0,2,2]$  
    \\
     &            &               &                     & 
$[-2,-2,-2,-2,0,0,2,2]$      \\
\hline
4    & --         & --            & --                  & 
$[-3,-3,-1,-1,1,1,3,3]$     \\
\hline
\end{tabular}
\end{center}
\caption{\small
Table of contributing cells up to 8 particles. For each 
class, we only give a cell of normal form generating them; 
the rest can be obtained by permuting the integers or changing their signs.}
\label{boxes}
\end{table}

\begin{figure}[h]
\begin{center}
\includegraphics[scale=0.7]{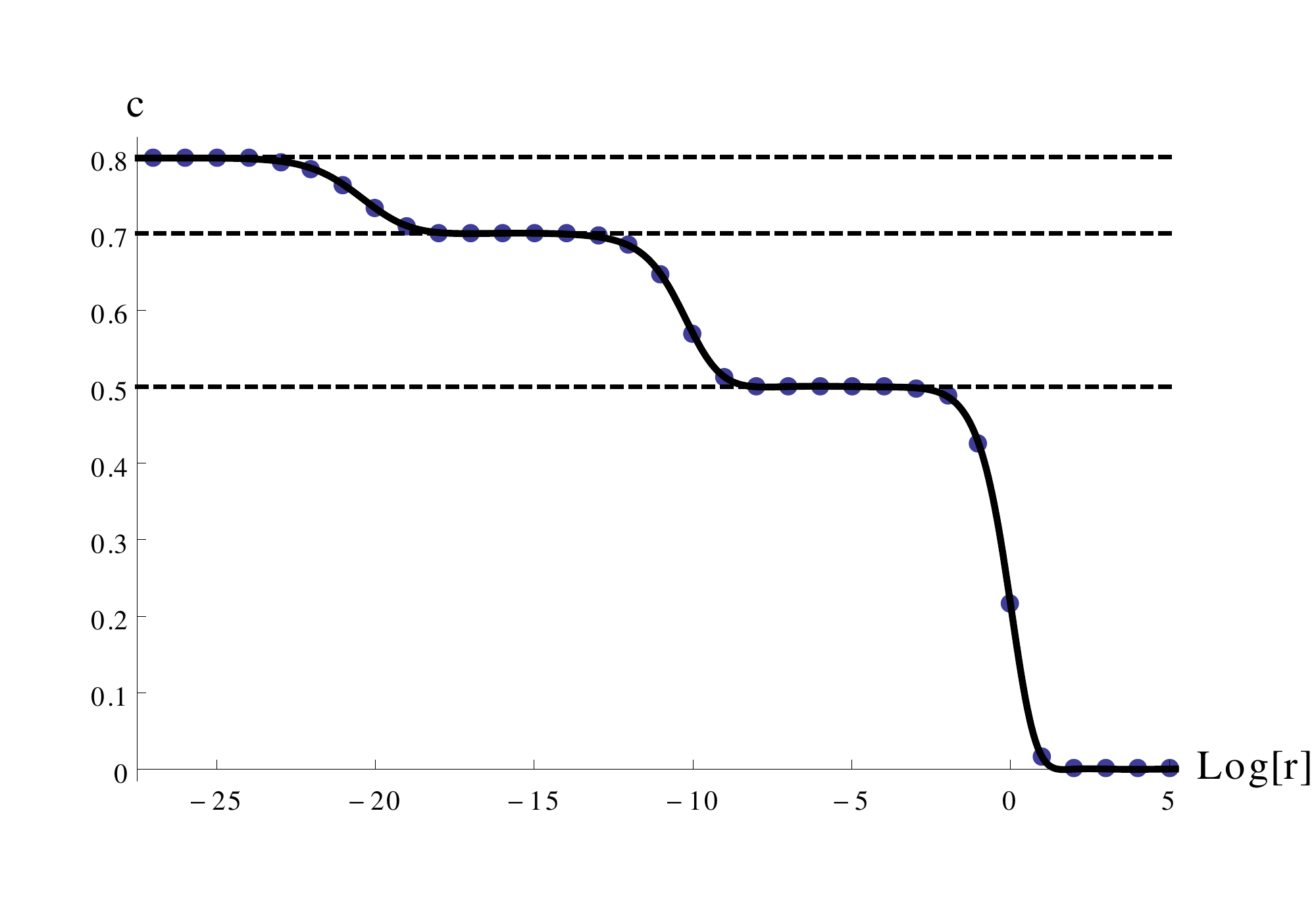} 
\end{center}
\caption{\small
The first three steps of the staircase for 
$\vartheta_0=20$, including 2,4,6 and 8 particle contributions. 
The dots show the result of numerical integration, the curve is a fit by an 
appropriate rational function of $r$ to 
the numerical integration results, while the horizontal lines show the central 
charges of the first three minimal models.}
\label{3steps}
\end{figure}

Using the localisation of the integration, it is possible to obtain a much more 
accurate result for the flow, as well as to include the third step 
as shown in figure \ref{3steps}. The plateau values which we obtain are $c=0.5$ 
(exact), $c=0.6999\dots$ and $c=0.799\dots$ where we terminate the 
results with the last accurate digit; these are all in good agreement with the 
minimal model predictions. For the fourth step, the 8-particle 
contribution (for $\vartheta_0=20$) is $\Delta c=0.0470\dots$, while the total 
expected difference between the corresponding minimal models is 
$\Delta c=4/70=0.05714\dots$. Again, the majority of the central charge 
difference arises at the first level that contributes to the given
step; however
reliable numerical evaluation of higher many-particle integrals proved too
difficult,
preventing the reconstruction of this step to better accuracy, 
even using the help of localisation.

The pattern in Table \ref{boxes} leads us to formulate the following 

\begin{paragraph}{Theorem R (relocalisation):}\label{theoremR}
The cells contributing to the $k$th step have the following normal form:
\begin{equation}
[a_1,a_1,\dots,a_p,a_p]\; :\; p\geq k\; ,\;-a_1=a_p=k-1\; ,\; 
a_k-a_{k-1}=0\;\mathrm{ or }\;2
 \label{guysrule}
\end{equation}
\end{paragraph}

In the following section we prove this statement using analytic 
considerations.

\resection{The form factor integrals in the staircase limit}
\label{analytic}

\subsection{Scaling with the distance parameter $r$}

To understand the behaviour of the $c$-function, we start by considering 
the explicit $r$-dependence of the integrand
in the c-theorem sum rule (\ref{c-sumrule}).

The factor
\begin{equation}
 e^{-rE}=\prod\limits_{i=1}^n e^{-r\cosh\theta_i}
\end{equation}
is the most dominant and the simplest to analyze. Each subfactor obeys
\begin{equation}
e^{-r\cosh\theta_i}=
\begin{cases}
1 & |\theta_i|\ll-\log r \\
0 & |\theta_i|\gg-\log r 
\end{cases}\; ,
\end{equation}
with the transitional region of $|\theta_i|$ 
around $-\log r$ having a thickness 
of $O(1)$, i.e. it does not scale with the relevant parameters $\log r$ 
or $\vartheta_0$. This means 
at a given value of $r$, the contributing region is essentially the hypercube
\begin{equation}
|\theta_i|\leq -\log r + O(1)\; ,
\end{equation}
where the $O(1)$ term 
indicates that the contribution decays in a double 
exponential way outside 
the hypercube. Note that all the other terms in the integrand (including the 
form factors) have at 
most exponential behaviour in rapidities, so they cannot counteract this 
behaviour. 

In conclusion, for the $k$th step located at
\begin{equation}
 -\log r=(k-1)\vartheta_0\; ,
\end{equation}
the above considerations imply that contributing cells have coordinates 
$[a_1,\dots,a_n]$ such that
\begin{equation}
 |a_i|\leq k-1\quad\mathrm{for }\;i=1,\dots,n\, .
\end{equation}

\subsection{Behaviour of the form factors}\label{sec:naive_FF_powercounting}

We now wish to examine the form factors of the trace of the
energy momentum tensor (\ref{eq:Theta0exp1}) in the roaming 
limit (\ref{eq:sinpibper2roaming}). 
Using the Lukyanov representation (\ref{eq:lukyanov_rep}) one obtains
\begin{align}
&P_{n}^{(1)}\left(\theta_{1},\dots,\theta_{n}\right)=\nn\\
&\qquad
\sum_{\{\alpha_{j}=\pm1\}}\left\{
\left(\prod_{j=1}^{n}\alpha_{j}e^{-\alpha_{j}(\vartheta_0+i\pi/2)}\right)
\prod_{r<s}\left(1-i\frac{\alpha_{r}{-}\alpha_{s}}{2}\frac{\cosh\vartheta_0}{
\sinh\left(\theta_{r}{-}\theta_{s}\right)}\right)\right\}\; .
\label{eq:lukpoly1}
\end{align}
To identify the relevant limiting behaviour we will need let
$\vartheta_0\to\infty$ in this expression
while simultaneously taking suitable rapidity
differences to infinity, so as to capture the relocalised
contributions from the various cells identified in theorem R of the 
previous section. These `double scaling' limits are significantly more
delicate than those leading to the cluster property 
(\ref{eq:ffclustering}) \cite{Delfino:1996nf}, in which certain
rapidity differences are taken to infinity at fixed coupling, and
their discussion will take up most of the rest of this section.

\subsubsection{Asymptotics of the minimal form
factor and normalisation constants}\label{sec:roamingminff}
As a preliminary step, we
need the asymptotic behaviour
of the minimal form factor, which can be written as
\begin{equation}
f(\theta)=\mathcal{N}R(\theta)\label{eq:fmin_NR}\; ,
\end{equation}
with
\begin{equation}
R(\theta)=\exp\left[8\int_{0}^{\infty}\frac{dt}{t}\sin^{2}\left(\frac{
t(i\pi-\theta)}{2\pi}\right)\frac{\sinh\frac{t\gamma}{2\pi}
\sinh(1-\frac{\gamma}{\pi})\frac{t}{
2}\sinh\frac{t}{2}}{\sinh^{2}t}\right]\; .
\label{eq:fmin_Rint}
\end{equation}
Explicit calculation shows that leading asymptotic behaviour
of this function is given by
\begin{eqnarray}
R(\theta+k\vartheta_0) &
\mathop{\longrightarrow}\limits_{\vartheta_0\rightarrow0} & \begin{cases}
-i\sinh\theta/2 & k=0\\
e^{\vartheta_0/2}\rho(\theta) & k=+1\\
e^{\vartheta_0/2}\bar{\rho}(\theta) & k=-1\\
\frac{1}{2}e^{\vartheta_0/2} & |k|>1
\end{cases}
\; ,
\label{eq:rfunctionroaming}
\end{eqnarray}
where
\begin{eqnarray}
\rho(\theta) & = &
\frac{1-i}{4}2^{1/4}e^{-K/\pi}e^{\theta/4}\exp\left(-\int_{0}^{\infty}\frac{dt}{
t}\frac{\sin^{2}\left(\frac{t(i\pi-\theta)}{2\pi}\right)}{\sinh
t\cosh\frac{t}{2}}\right)\; ,\nonumber \\
\bar{\rho}(\theta) & = &
\frac{1+i}{4}2^{1/4}e^{-K/\pi}e^{-\theta/4}\exp\left(-\int_{0}^{\infty}\frac{dt}
{t}\frac{\sin^{2}\left(\frac{t(i\pi-\theta)}{2\pi}\right)}{\sinh
t\cosh\frac{t}{2}}\right)\; ,
\label{eq:rhofunction}
\end{eqnarray}
while
\begin{equation}
K=0.915966\dots\label{eq:Catalan}
\end{equation}
denotes Catalan's constant, and
\begin{align}
\rho(\infty)=\bar{\rho}(-\infty) & =  \frac{1}{2}\nonumber \\
\rho(-\infty)=\bar{\rho}(\infty) & =  0\label{eq:rho_limits}\; .
\end{align}
In addition, the normalisation $\mathcal{N}$ factor behaves as
\begin{eqnarray}
\mathcal{N} & \sim & 2e^{-\vartheta_0/2}\; ,
\label{eq:calNroaming}
\end{eqnarray}
and consequently
\begin{equation}
F_{2}^{(1)}(0,i\pi)\sim-2e^{\vartheta_0}\; .
\label{eq:F2ipi}
\end{equation}

\subsubsection{Naive power counting}

Let us write
\begin{equation}
\theta_{i}=\theta_{i}'+t_{i}\vartheta_0\; ,
\label{eq:theta_finite_T0}
\end{equation}
where $\left|\theta_{i}'\right|$ is kept finite as
$\vartheta_0\rightarrow\infty$. This corresponds to the observation in section 
\ref{cellsandloc} that the rapidity integration domain can be
split into cells $[a_{1},\dots,a_{n}]$. Due to Lorentz invariance (\ref{eq:Lorentzaxiom}), 
the form factors only depend on rapidity differences, and according to the form factor equation 
(\ref{eq:exchangeaxiom}), exchanging two rapidities gives just a phase factor. 
Therefore for the form factor part of the discussion we can perform a
global shift and reshuffling of the rapdities to render
the $t_{i}$ ordered and positive, with $t_1=1$:
\begin{equation}
1=t_{1}\leq\dots\leq t_{n}\; .\label{eq:t_ordered}
\end{equation}
For a cell of normal form $[a_{1},\dots,a_{n}]$ the 
corresponding $\vec{t}$ is given by
\begin{equation}
t_i=(a_i-a_1)/2+1\quad i=1,\dots,n\; ;
\end{equation}
we will refer to such a cell as being of type $\vec{t}$.
Furthermore, the asymptotic behaviour depends only on
\begin{equation}
t_{ij}=t_{i}-t_{j}=(a_i-a_j)/2\; .\label{eq:deltat}
\end{equation}
Since our numerical studies showed that the 
cell coordinates $[a_{1},\dots,a_{n}]$ 
of contributing cells are integers of the same parity, we now make the following
\begin{paragraph}
{Assumption I (integrality):}\label{assumptionI}
 all the $t_i$ are integers. Since $t_1$ is fixed to $1$, this is equivalent to
\begin{equation}
t_{ij}\in\mathbb{Z}\; .
 \label{integrality}
\end{equation} 
\end{paragraph}

In Section \ref{sec:integrality} the validity of this assumption will be 
demonstrated by analytic considerations.

The large-$\vartheta_0$ asymptotic  behaviour of the form factors in a
cell of type  $\vec{t}$  
is characterised by an exponent $\omega(\vec{t}\:)$ defined as
\begin{equation}
F_{n}^{\Theta}\left(\theta_{1},\dots,\theta_{n}\right)\sim
e^{\omega(\vec{t}\:)\vartheta_0}\label{eq:omegatdef}
\end{equation}
or, more precisely
\begin{equation}
\omega(\vec{t}\:):=\lim\limits_{\vartheta_0\rightarrow\infty} \frac{1}{\vartheta_0}\log 
F_{n}^{\Theta}\left(\theta_{1}'+t_1\vartheta_0,\dots,\theta_{n}'+t_n\vartheta_0\right)\; .
\end{equation}
Using (\ref{eq:lukyanov_rep}), (\ref{eq:rfunctionroaming}) and
(\ref{eq:F2ipi}) one can write
\begin{equation}
\omega(\vec{t}\:)=-1-\frac{n}{4}+\omega_{P}(\vec{t}\:)+\omega_{F}(\vec{t}\:)\; ,
\label{eq:omegatsplit}
\end{equation}
where the constant $-1$ comes from normalizing by $F_{2}(0,i\pi)$,
the term $\omega_{F}(\vec{t}\:)$ is the contribution of the minimal
form factors, which can be written as
\begin{equation}
\omega_{F}(\vec{t}\:)=\sum_{<i,j>}\min\left(0,\frac{|t_{ij}|-1}{2}\right)
\; ,
\label{eq:omegaFt}
\end{equation}
while $\omega_{P}(\vec{t}\:)$ is the asymptotic behaviour of the function 
$P^{(1)}_n$ 
defined in (\ref{eq:lukpoly1}).

\subsubsection{Upper estimate for $\omega_{P}$}\label{sec:oP_upper_estimate}

The form factor polynomial $P_{n}^{(1)}$ (\ref{eq:lukpoly1}) is
the sum of $2^{n}$ terms of the form
\begin{equation}
\left(\prod_{j=1}^{n}\alpha_{j}e^{-\alpha_{j}(\vartheta_0+i\pi/2)}\right)\prod_{
k<j}\left(1-i\frac{\alpha_{k}-\alpha_{j}}{2}\frac{\cosh\vartheta_0}{
\sinh\left(\theta_{k}-\theta_{j}\right)}\right)\sim
e^{\omega_{P}(\vec{\alpha},\vec{t}\:)\vartheta_0}\; ,
\label{eq:P1nterms}
\end{equation}
where $\omega_{P}(\vec{\alpha},\vec{t}\:)$ is the exponent
characterizing
the asymptotic behaviour of these terms for each value of
$\vec{\alpha}$. Then the following inequality
\begin{equation}
\omega_{P}(\vec{t}\:)\leq\tilde{\omega}_{P}(\vec{t}\:)=\max_{\vec{\alpha}}\omega_{P}
(\vec{\alpha},\vec{t}\:)\label{eq:omegaP_maxalpha}
\end{equation}
provides an upper estimate for the asymptotic behaviour. 

To analyze the $\vec{\alpha}$ dependence we can start with the case
\begin{equation}
\alpha_{1}=\dots=\alpha_{n}=-1\; ,
\label{eq:alpha_ref}
\end{equation}
for which we obtain
\begin{equation}
\tilde{\omega}_{P}(\vec{\alpha},\vec{t}\:)=n\; .
\label{eq:alpha_ref_power}
\end{equation}
We can then consider the effect of flipping some of the $\alpha_k$ from  $-1$ to $+1$.
It is convenient to introduce a notation in which $\vec{t}$ is parameterised
by $N$ blocks of lengths $p_k$, $k=1,\dots,N$ such that in the $k$th block 
the components of $\vec{t}$ are equal to a constant value $v_k$:
\begin{equation}
\vec{t}=(\underbrace{v_{1}\dots v_{1}}_{p_{1}},\underbrace{v_{2}\dots
v_{2}}_{p_{2}},\dots,\underbrace{v_{N}\dots
v_{N}}_{p_{N}})\; ,
\label{eq:blocksdef}
\end{equation}
where $v_{N}=\max\vec{t}$, $v_{1}<v_{2}<\dots<v_{N}$ and
\begin{equation}
\sum_{k=1}^{N}p_{k}=n\; .\label{eq:sumblocks}
\end{equation}
Let us consider the case when there are $r_{1}$ instances of
$\vec{\alpha}$
components equal to $+1$ in the first block, $r_{2}$ in the second
etc. In terms of $\tilde{\omega}_{P}(\vec{\alpha},\vec{t}\:)$, each such 
flip comes with a ``cost'' of $-2$ from the prefactor
in the $\vec{\alpha}$-term, while flips in the same block also make
``gains'' of $+1$ for by ``activating'' some of the $\cosh\vartheta_0$
terms. The number of such activated terms is given by the number of
$+1,-1$ pairs inside the blocks, so we obtain
\begin{equation}
\tilde{\omega}_{P}(\vec{\alpha},\vec{t}\:)=n+\sum_{k=1}^{N}(-2r_{k}+(p_{k}-r_{k}
)r_{k})\; .
\label{eq:blockpowerestimate}
\end{equation}
Since the contributions of the blocks are independent, we can maximise
the terms separately:
\begin{equation}
-2r_{k}+(p_{k}-r_{k})r_{k}\quad\mbox{ is maximal when }\quad
r_{k}=\left[\frac{p_{k}-2}{2}\right]\; .
\label{eq:blockpowermaximize}
\end{equation}
The end result is 
\begin{eqnarray}
\tilde{\omega}_{P}(\vec{t}\:)&=&
n+\sum_{k=1}^{N}\left[\left(\frac{p_{k}-2}{2}\right)^{2}
-\frac{\delta_k^2}{4}\right]
\nonumber\\
&&
\delta_k=
\begin{cases}
0 & p_k\;\mathrm{even}\\
1 & p_k\;\mathrm{odd}
\end{cases}\; .
\label{eq:naiveomegaP}
\end{eqnarray}

\subsubsection{Upper estimate for $\omega$ and dominant cells}

Substituting (\ref{eq:naiveomegaP}) and (\ref{eq:omegaFt}) into (\ref{eq:omegatsplit}), 
the end result for the upper estimate $\tilde\omega$ is
\begin{equation}
\omega(\vec{t}\:)\leq\tilde{\omega}(\vec{t}\:)=-1+\frac{3n}{4}+\sum_{k=1}^{N}
\left(\frac{p_{k}-2}{2}\right)^{2}
-\sum_{k=1}^{N}\frac{\delta_k^2}{4}
+\sum_{\left\langle i,j\right\rangle
}\min\left(0,\frac{|t_{ij}|-1}{2}\right)\; ,
\label{eq:omegaestimate}
\end{equation}
where $\left\langle i,j\right\rangle $ denotes an (unordered) pair
of indices between $1$ and $n$. Using the integrality assumption 
$t_{ij}\in\mathbb{Z}$ the last term can be computed explicitly,
since the summands are all $0$ except when $t_{ij}=0$, which is
when $\left\langle i,j\right\rangle $ is a pair from inside the same
block. Such pairs contribute $-1/2$, so
\begin{eqnarray}
\tilde{\omega}(\vec{t}\:) & = &
-1+\frac{3n}{4}+\sum_{k=1}^{N}\left[\left(\frac{p_{k}-2}{2}\right)^{2}-\frac{p_{
k}(p_{k}-1)}{4}\right]-\sum_{k=1}^{N}\frac{\delta_k^2}{4}\nonumber
\\
 & = &
-1+\sum_{k=1}^{N}\left(\frac{3}{4}p_{k}+\frac{4-3p_{k}}{4}\right)-\sum_{k=1}^{N}\frac{\delta_k^2}{4}\nonumber
\\
 & = & -1+N-\sum_{k=1}^{N}\frac{\delta_k^2}{4}\; .\label{eq:omegaestimatesimplified}
\end{eqnarray}
This is surprisingly simple: the end result is just the number of
blocks minus 1, with a negative correction for odd blocks. 

In the following section we show that the relevant issue is to find the dominant blocks for 
a fixed value of 
\begin{equation}
D(\vec{t}\:)=\max\vec{t}-\min\vec{t} \; ,
\label{eq:Dt_def}
\end{equation}
which measures the real space ``length'' of the sequence $\vec{t}$. In this case the way 
to maximise the number of blocks is to minimise the steps between blocks, so for the dominant 
cells integers in adjacent blocks differ by $1$:
\begin{equation}
 v_{k+1}-v_k=1\; . 
\end{equation}
The total number of blocks is then given by 
$N=D(\vec{t}\:)+1$. To further maximise the contribution all block 
lengths should be chosen even. As a result, for any fixed value of $D(\vec{t}\:)$ 
the dominant contributions comes from cells which have the largest possible number of blocks, 
all of which are even. 

Note that so far we have only obtained an upper limit for the scaling exponent $\omega(\vec{t}\:)$. 
This can be taken into account by writing the true exponent in the form
\begin{equation}
\omega(\vec{t}\:)=\tilde{\omega}(\vec{t}\:)+\Delta\omega(\vec{t}\:)\; ,
\label{eq:anomalydef}
\end{equation}
where
\begin{equation}
\Delta\omega(\vec{t}\:)=\omega_{P}(\vec{t}\:)-\tilde{\omega}_{P}(\vec{t}\:)\leq 0
\end{equation}
will be called the anomaly term. This arises from potential cancellations
between terms with different $\vec{\alpha}$, to which we return later
in Section \ref{sec:anomaly}.

\subsection{$c$-function in the roaming limit: proof of Theorem \textbf{R}}
\label{sec:cfunction_scaling}

Now let us examine the $c$-function contributions. Let us take a cell 
$[a_1,\dots,a_n]$ of normal form, i.e.
\begin{equation}
a_1\leq\dots\leq a_n\; ,
\end{equation}
and recall that the corresponding $t$ sequence is
\begin{equation}
t_i=(a_i-a_1)/2+1\quad i=1,\dots,n\; .
\end{equation}
The upper limit for the form factor exponent is given by
\begin{equation}
 {\omega}(\vec{t}\:)\leq \tilde{\omega}(\vec{t}\:)= N-1\; ,
\end{equation}
where $N$ is the number of blocks in $\vec{t}$. For a given cell
to contribute to $c(r)$ its coordinates must satisfy
\begin{equation}
 \log r\leq a_1\leq\dots\leq a_n\leq -\log r \; .
\label{eq:logr_cellcondition}
\end{equation}
The contribution of a cell $[\vec{a}]$ satisfying (\ref{eq:logr_cellcondition}) 
to the $c$-function (\ref{c-sumrule}) is given by  
\begin{equation}
\int_{[\vec{a}]}\!\frac{d\theta_1\dots d\theta_{n}}
{(2\pi)^n(n!)}\,
\frac{6+6rE+3r^2E^2+r^3E^3}{2E^4}\,
|m^{-2} F_{n}^{\Theta}(\theta_1\dots\theta_{n})|^2\,e^{-rE} \; .
\end{equation}
For the values of $r$ when the given cell contributes $rE$ is $O(1)$, i.e. it is bounded by a $\vartheta_0$ 
independent constant when scaling $\vartheta_0\rightarrow\infty$ (in fact $rE\lesssim n$). 
Therefore the scaling power $\zeta(\vec{a})$ of the $c$-function 
contribution from cell $[\vec{a}]$ can be defined as the behaviour of the 
integral
\begin{equation}
\int_{[\vec{a}]}\!\frac{d\theta_1\dots d\theta_{n}}
{(2\pi)^n(n!)}\,
\frac{|m^{-2} F_{n}^{\Theta}(\theta_1\dots\theta_{n})|^2}{2E^4}\propto 
e^{\zeta(\vec{a}) \vartheta_0}
\end{equation}
or, more precisely
\begin{equation}
\zeta(\vec{a})
:=\mathop{\mathrm{lim}}\limits_{\vartheta_0\rightarrow\infty}\;\frac{1}{
\vartheta_0}\log
\int_{[\vec{a}]}\!\frac{d\theta_1\dots d\theta_{n}}
{(2\pi)^n(n!)}\,
\frac{|m^{-2} F_{n}^{\Theta}(\theta_1\dots\theta_{n})|^2}{2E^4}\; .
\end{equation}
The behaviour of the form factor depends only the differences $t_{ij}=t_i-t_j$; 
however the energy denominator breaks translational invariance in rapidity space. For a given $\vec{t}$ the value 
of $a_n-a_1=2 (D(\vec{t}\:)-1)$ is fixed, while $a_1$ is fixed by the energy denominator, which is 
minimised when $\mathop{\mathrm max}\limits_i |a_i|=a_n$ takes the smallest possible value. As a result 
the maximum contribution is obtained when $a_1=-a_n$, all other cases being suppressed by powers 
of $e^{\vartheta_0}$. Therefore the cells giving the highest possible contribution
have the form
\begin{equation}
[\vec{a}]=[\underbrace{-k+1,\dots,-k+1}_{p_1},
\underbrace{-k+1,\dots,-k+1}_{p_{2}},\dots,
\underbrace{k-1,\dots,k-1}_{p_{k}}]\; ,
\label{eq:guysrule}
\end{equation}
where $k=D(\vec{t}\:)$, each $p_{l}$ is a positive even number, and $(k-1)\vartheta_0/2\leq-\log r$. 
For such a cell the number of blocks is exactly $N=k$ and the energy 
denominator has the behaviour 
\begin{equation}
 E^4\propto \exp\left\{ 2 (k-1)\vartheta_0\right\}\; ,
\end{equation}
so that the contribution of a cell (\ref{eq:guysrule}) to the $c$-function scale with the exponent 
\begin{equation}
\zeta(\vec{a}) = 2\omega(\vec{t}\:)-2(k-1) < 
2(N-k)=0\; .
\label{eq:stepk_cell_naivepover}
\end{equation}
This shows that the contribution of dominant cells scales to at most a constant when 
$\vartheta_0\rightarrow \infty$. Therefore none of the subdominant cells can give any contributions 
in this limit. More precisely, since our estimate is only an upper one, it can be stated that 
cells (\ref{eq:guysrule}) are the only ones which have any chance at
all to
give a  non-vanishing contributions in the roaming limit.
However, at this stage we cannot be certain whether all of them contribute. In Section 
\ref{sec:anomaly} below we show that for these cells the anomaly term vanishes, 
therefore they all contribute in the roaming limit, with the exception of the $k=1$ case, where the 
only contributing cell is $[0,0]$. 

Our result also explains the existence of steps in $c(r)$. Since (under the assumption of integrality) 
$k$ can only take positive  integer values, in the $O(1)$ vicinity of the special values
\begin{equation}
 \log r=-\frac{k-1}{2}\vartheta_0\; , \quad k=1,2,\dots
\end{equation}
new cells are switched on, giving rise to the staircase pattern in $c(r)$, 
with crossovers in the vicinity of the above values of $\log r$, and plateau 
connecting the crossover regions to each other.

The above argument proves Theorem \textbf{R} stated in section \ref{cellsandloc}, provided the 
integrality assumption $t_{ij}\in\mathbb{Z}$ is taken granted.

\subsection{The anomaly term}\label{sec:anomaly}

Recall that so far we only established an upper bound on cell contributions. 
By considering the scaling anomaly defined in (\ref{eq:anomalydef}) this can be 
refined into an exact result. The anomaly term results from cancellations 
between terms with different
$\vec{\alpha}$ in (\ref{eq:lukpoly1}), which reduce the exponent
of $e^{\vartheta_0}$ compared to the naive power counting performed
so far. Suppose $\vec{t}$ has the form
\begin{eqnarray}
\vec{t} & = & (\underbrace{v_{1}\dots
v_{1}}_{p_{1}},\underbrace{v_{2}\dots
v_{2}}_{p_{2}},\dots,\underbrace{v_{N}\dots v_{N}}_{p_{N}})\nonumber\\
 &  & 1= v_{1}<v_{2}<\dots<v_{N}\leq n
\end{eqnarray}
Let us introduce the following classification: an index pair
$\left\langle j,k\right\rangle $
is called
\begin{enumerate}
\item \emph{internal} if $|t_{jk}|=0$; this means $j$ and $k$ have
$a_{j}=a_{k}$,
i.e. they are from inside the same block;
\item \emph{strongly linked} if $|t_{jk}|=1$;
\item \emph{weakly linked} if $|t_{jk}|\geq2$.
\end{enumerate}

The vector $\vec{t}$ can be partitioned into clusters by weak links.
A cluster consists of blocks which are separated by strong links;
a cluster is called \emph{primitive} if it consists of a single block.
In (\ref{eq:P1nterms}) weak links contribute only a factor $1$ in the limit 
$\vartheta_0\rightarrow\infty$, therefore
\begin{align}
&P_{n}^{(1)}\left(\theta_{1},\dots,\theta_{n}\right)\sim \\
&~~~
\prod_{\mbox{\small clusters}\,\mathcal{C}}\,
\sum_{\vec{\alpha}\in A(\mathcal{C})} \!
\left\{
\!\left(\prod_{j\in\mathcal{C}}\alpha_{j}
e^{-\alpha_{j}(\vartheta_0+i\pi/2)}\!\right)\!
\prod_{<j,k>\in\mathcal{C}}\left[\left(1-i\frac{\alpha_{k}{-}\alpha_{j}}{2}\frac
{\cosh\vartheta_0}{\sinh\left(\theta_{k}{-}\theta_{j}\right)}
\right)\right]\right\}\; ,\nonumber
\end{align}
where $A(\mathcal{C})$ is the set of $\vec{\alpha}$ configurations
inside $\mathcal{C}$. Due to this factorisation property cancellations 
can only happen within clusters.

We digress a little to discuss the relation of this classification with the 
form factor cluster property (\ref{eq:Thetaclustering}). In the roaming 
case (\ref{eq:sinpibper2roaming}) it can be rewritten as
\begin{eqnarray}
F_{n+m}^{\Theta}\left(\theta_{1}+\Lambda,\dots,\theta_{n}+\Lambda,
\theta_{1}',\dots,\theta_{m}'\right)
&=& \frac{2\cosh\vartheta_0}{\pi}
F_{n}^{\Theta}\left(\theta_{1},\dots,\theta_{n}\right)
F_{m}^{\Theta}\left(\theta_{1}',\dots,\theta_{m}'\right) \nonumber\\
& &+O\left(e^{-\Lambda}\right).
\label{eq:roamingThetaclustering}
\end{eqnarray}
For $\vartheta_0$ finite this leads to a twofold classification of links into
internal and weak ones, whereas internal links are those between rapidities separated 
by finite distance as $\Lambda\rightarrow\infty$, while weak links are those which are between 
rapidities whose distance grows with $\Lambda$. However, in our case the classification 
is more refined as $\vartheta_0\rightarrow\infty$. This results from the fact that 
\begin{equation}
\mathrm{for }\quad |\theta_k-\theta_j|=\alpha\vartheta_0+O(1):\quad 
\frac{\cosh\vartheta_0}{\sinh\left(\theta_{k}{-}\theta_{j}\right)}\rightarrow 
\begin{cases}
\infty & \alpha<1 \\
O(1) & \alpha=1 \\
0    & \alpha>1 
\end{cases}\; ,
\end{equation}
which exactly corresponds to the threefold (internal/strong/weak) classification introduced above.

Returning to the anomaly, let us first consider a primitive cluster 
$\bar{\mathcal{C}}$ of length $p$; the factor corresponding to cluster 
$\bar{\mathcal{C}}$ is exactly the $p$-particle form factor polynomial 
in the variables $\theta_{j}$ with $j\in\bar{\mathcal{C}}$, which 
contributes the power
\begin{equation}
\omega_{\bar{\mathcal{C}}}=\omega_{P}(\underbrace{1\dots1}_{p})\; .
\end{equation}
On the other hand from (\ref{eq:km_lukyrel})
\begin{equation}
\omega_{P}(\underbrace{1\dots1}_{p})=\omega_{Q}(p)+p \; ,
\end{equation}
where $\omega_{Q}(p)$ is defined by the asymptotics
\begin{equation}
Q_{p}(\theta_{1},\dots\theta_{p})\sim e^{\omega_{Q}(p)\vartheta_0}
\end{equation}
for $\vartheta_0\rightarrow\infty$ with $\theta_{i}$ kept finite.
The recursion relation (\ref{eq:Qrecursion}) gives
\begin{equation}
\omega_{Q}(p+2)=\omega_{Q}(p)+p-2\; ,
\end{equation}
and using the value $\omega_{Q}(2)=0$ we obtain
\begin{equation}
\omega_{Q}(p)=\frac{(p-2)^{2}}{4}-\frac{p-2}{2}
\end{equation}
resulting in
\begin{equation}
\omega_{P}(\underbrace{1\dots1}_{p})=\frac{(p-2)^{2}}{4}+\frac{p+2}{2}\; .
\end{equation}
The naive value is given by the $N=1$ case of (\ref{eq:naiveomegaP})
\begin{equation}
\tilde{\omega}_{P}(\underbrace{1\dots1}_{p})=p+\frac{(p-2)^{2}}{4}\; .
\end{equation}
so the anomaly for a single primitive cluster is
\begin{equation}
\Delta\omega(\underbrace{1\dots1}_{p})=\omega_{P}(\underbrace{1\dots1}_{p}
)-\tilde{\omega}_{P}(\underbrace{1\dots1}_{p})=-\frac{p-2}{2}\; .
\end{equation}
Non-primitive clusters are composed of several blocks connected by
strong links, which are of order $1$ and depend on relative
rapidities between the clusters, which can take any finite value.
As a result, they upset the cancellation between terms with different
$\vec{\alpha}$ which is responsible for the anomaly. Therefore only
primitive clusters contribute to the anomaly, and their contributions
are additive due to the factorisation between clusters. 

Our final result for the anomaly is
\begin{equation}
\Delta\omega(\vec{t}\:)=~-\!\!\!\!\!\!\sum_{{
\genfrac{}{}{0pt}{}{\mathcal{C}\mbox{\tiny : primitive}}{\mbox{\tiny cluster}}
}}\!\!\!\frac{\mbox{length}(\mathcal{C})-2}{2}\; .
\end{equation}

Let us examine the sequences
\begin{equation}
\vec{t}=(\underbrace{1\dots1}_{p_{1}},\underbrace{2\dots2}_{p_{2}},\dots,
\underbrace{k\dots k}_{p_{k}})\; ,
\end{equation}
which the naive power counting identified as the ones potentially
contributing to the $c$-function at step $k$. For $k\geq2$ $\vec{t}$
is composed of a single non-primitive cluster, therefore the anomaly
is zero. At step $k=1$ the naively contributing cells form a primitive
cluster
\begin{equation}
\vec{t}=(\underbrace{1\dots1}_{p})\; ,
\end{equation}
and their $e^{\vartheta_0}$ power in the $c$-function is
\begin{equation}
\zeta(\vec{t}\:)  = 
2\omega(\vec{t}\:)-2(k-1)=2\Delta\omega(\vec{t}\:)=-p+2\; .
\end{equation}
This agrees with the fact that the first step (corresponding to the
Ising model) receives only two-particle ($p=2$) contributions.

\subsection{Eliminating the integrality assumption}\label{sec:integrality}

We recall that in the derivation of the contributing cells, our crucial starting 
assumption (inferred from the numerics) was that we can parameterise the relevant 
regions in the rapidity space as 
\begin{equation}
\theta_{i}=\theta_{i}'+t_{i}\vartheta_0\; ,
\end{equation}
where 
\begin{equation}
t_{ij}=t_i-t_j
\end{equation}
are integers, while $\theta_{i}'$ are finite while 
$\vartheta_0\rightarrow\infty$.
Because of this limit, however, vast parts of the rapidity space are not 
covered by the arguments. Here we argue that these domains do not contribute to tha 
$c$-function (\ref{c-sumrule}) in the limit $\vartheta_0\rightarrow \infty$. 

To begin with, we note that the cells satisfying the rule formulated in Theorem R
all contribute a finite amount in the limit $\vartheta_0\rightarrow \infty$. Therefore to 
eliminate any other possibilities it is enough to demonstrate their suppression 
relative to these cells by a factor which vanishes in the same limit. In fact, what 
can be demonstrated is that relaxing assumption I introduces 
a suppression which is exponential in $\vartheta_0$. 

Following the conventions introduced in Section \ref{sec:oP_upper_estimate} we use the 
block notation
\begin{eqnarray}
\vec{t} & = & (\underbrace{v_{1}\dots v_{1}}_{p_{1}},\underbrace{v_{2}\dots 
v_{2}}_{p_{2}},\dots,\underbrace{v_{N}\dots v_{N}}_{p_{N}})\nonumber\\
 &  & v_{1}<v_{2}<\dots<v_{N}\nonumber\\
 &  & t_{ij}=t_{i}-t_{j}\qquad v_{ij}=v_{i}-v_{j}\;,
\end{eqnarray}
and also recall the definition of  the ``length'' of $\vec{t}$: 
\begin{equation}
 D(\vec{t}\:)  =  v_{N}-v_{1}\;.
\end{equation}
The results (\ref{eq:omegaFt}) and (\ref{eq:blockpowerestimate}) can be 
rewritten to take into account that the $t_{ij}$ are not necessarily integers. 
The contribution (\ref{eq:omegaFt}) of the minimal form factors can be written as
\begin{equation}
\omega_{F}(\vec{\alpha},\vec{t}\:)=-\sum_{k=1}^{N}\frac{p_{k}(p_{k}-1)}{4}+\sum_{
<i,j>:|v_{ij}|<1}p_{i}p_{j}\frac{|v_{ij}|-1}{2}\; ,
\end{equation}
where the last term is a correction by the minimal form factors for which $t_{ij}<1$. 
The generalisation of (\ref{eq:blockpowerestimate}) is
\begin{equation}
\tilde{\omega}_{P}(\vec{\alpha},\vec{t}\:)=n+\sum_{k=1}^{N}(-2r_{k}+(p_{k}-r_{k}
)r_{k})+\sum_{<i,j>:|v_{ij}|<1}\left(r_{i}(p_{j}-r_{j})+r_{j}(p_{i}-r_{i}
)\right)(1-|v_{ij}|)\; ,
\end{equation}
where the last term is contributed by inter-block links that are ``activated''
by the choice of $\vec{\alpha}$. 
Substituting these contributions into (\ref{eq:omegatsplit}) results in 
\begin{eqnarray}
\tilde{\omega}(\vec{\alpha},\vec{t}\:)=-1+\frac{3n}{4}+\sum_{k=1}^{N}(-2r_{k}+(p_{
k}-r_{k})r_{k})-\sum_{k=1}^{N}\frac{p_{k}(p_{k}-1)}{4}\nonumber\\
-\sum_{<i,j>}\frac{1-|v_{ij}|}{2}(p_{i}-2r_{i})(p_{j}-2r_{j})\theta(1-|v_{ij}|)
\; .
\end{eqnarray}
Introducing the notation $s_{k}=p_{k}-2r_{k}$:
\begin{eqnarray}
\tilde{\omega}(\vec{\alpha},\vec{t}\:) & = & 
-1+\frac{3n}{4}+\frac{1}{4}\sum_{k=1}^{N}p_{k}^{2}-4p_{k}-\frac{1}{4}\sum_{k=1}^
{N}\left(s_{k}^{2}-4s_{k}\right)-\sum_{k=1}^{N}\frac{p_{k}(p_{k}-1)}{4}-\sum_{<i
,j>}\frac{A_{ij}}{2}s_{i}s_{j}\nonumber\\
 & = & 
-1+N-\frac{1}{4}\sum_{k=1}^{N}\left(s_{k}-2\right)^{2}-\sum_{<i,j>}\frac{A_{ij}}
{2}s_{i}s_{j} \; ,
\label{eq:omega_s}\end{eqnarray}
where the matrix $A$ is defined as
\begin{equation}
A_{ij}=\frac{1-|v_{ij}|}{2}\theta(1-|v_{ij}|) \; .
\end{equation}
It is important to note that $0\leq A_{ij}<1/2$ for all $i,j$. The
maximum value of $\tilde{\omega}$ is determined by the condition
\begin{equation}
-\frac{1}{2}s_{k}+1-\sum_{j}A_{kj}s_{j}=0 \; ,
\end{equation}
which can be written in matrix notation as
\begin{eqnarray}
(2A+1)\vec{s} & = & \vec{\rho}\nonumber\\
\mbox{where} &  & \vec{\rho}=(2,2,\dots,2) \; .
\end{eqnarray}
Note that $s_{k}$ must be integer with the same parity as $p_{k}$;
if any of the $s_{k}$ is not then it must be ``rounded'' to one of 
the nearest such values. However, such a rounding can only decrease 
$\tilde{\omega}(\vec{t}\:)$ so for an upper estimate it can be neglected,
which often simplifies the arguments. 

Let us first assume that $A=0$. Then the solution is $s_{k}=2$, which gives
\begin{equation}
\tilde{\omega}(\vec{t}\:)=-1+N \; .
\end{equation}
However, this value is only allowed when all the $p_{k}$ are even,
since for odd $p_{k}$ only odd $s_{k}$ is possible. As a result
the rounded values for $\vec{s}$ are
\begin{eqnarray}
\vec{s} & = & \vec{\rho}+\vec{\delta}\nonumber \\
 &  & \delta_{k}=\begin{cases}
0 & p_{k}\mbox{ even}\\
\pm1 & p_{k}\mbox{ odd}
\end{cases} \; .
\label{eq:A=00003D0solution}
\end{eqnarray}
leading to
\begin{equation}
\tilde{\omega}(\vec{t}\:)=-1+N-\frac{1}{4}\sum_{k=1}^{N}\delta_{k}^{2} \; ,
\end{equation}
which coincides with our previous result (\ref{eq:omegaestimatesimplified}). 
Using the same argument for cell centring
as in Section \ref{sec:cfunction_scaling}, the $c$-function exponent is 
\begin{eqnarray}
\zeta(\vec{a}) & \leq & 2(\tilde{\omega}(\vec{t}\:)-D(\vec{t}\:))\nonumber\\
\mbox{where} &  & a_{k}=t_{k}-t_{1}-\frac{D(\vec{t}\:)}{2} \; .
\end{eqnarray}
In order to maximise this it is necessary that $\delta_{k}=0$, i.e. all
blocks must be even and to minimise $D(\vec{t}\:)$ with the assumption $A=0$ 
one must have
\begin{equation}
v_{k+1}-v_{k}=1 \; ,
\end{equation}
so our original results are recovered. 

In fact, even for the case $A\neq 0$, if there is any gap $v_{k+1}-v_{k}$ 
between subsequent blocks
that is larger than $1$ then it can always be decreased to $1$
without changing $\tilde{\omega}(\vec{t}\:)$, but gaining in the final
exponent $\zeta$ by decreasing $D(\vec{t}\:)$, i.e. by ``shortening''. 
So we can proceed by assuming that 
\begin{equation}
\left|v_{k+1}-v_{k}\right|\leq 1 \mbox{ for all } k \; .
\end{equation}
For the general case $A\neq 0$, the solution for $\vec{s}$ is 
\begin{equation}
\vec{s}_{*}  =  (1+2A)^{-1}\vec{\rho}
\end{equation}
provided $(1+2A)$ is invertible, and the maximum value of $\tilde{\omega}$
is given by 
\begin{eqnarray}
\tilde{\omega}(\vec{t}\:) & = & 
-1+N-\frac{1}{4}(\vec{s}_{*}-\vec{\rho})\cdot(\vec{s}_{*}-\vec{\rho})-\frac{1}{2
}\vec{s}_{*}A\vec{s}_{*}\nonumber\\
 & = & -1+N-\frac{1}{2}\vec{\rho}\frac{A}{1+2A}\vec{\rho}\nonumber\\
 & = & -1+N-\frac{1}{2}\vec{\rho}A\vec{s}_{*} \; .
\end{eqnarray}

\subsubsection{Maximum shortening}

Our formalism can be checked by looking at the case of maximum shortening:
\begin{equation}
|v_{k+1}-v_{k}|=0 \quad \forall\; k\; ,
\end{equation}
i.e. the blocks are eventually joined into a large single
one with the length being minimal i.e. $D(\vec{t}\:)=0$. In this case
\begin{equation}
A_{ij}=(1-\delta_{ij})/2 \; .
\end{equation}
In this case the solution for $\vec{s}\;$ is not unique due to the fact that $1+2A$
is a singular matrix: any vector whose elements sum to $2$ works
\begin{equation}
\sum_{i=1}^{N}s_{i}=2 \; .
\end{equation}
This is in accordance with the overall picture, as for the single
large block there is a single $s$ parameter, which is exactly 
\begin{equation}
s=\sum_{i=1}^{N}s_{i}
\end{equation}
and must be equal to $2$. One can explicitly evaluate
\begin{eqnarray}
\tilde{\omega} & = & -1+N-\frac{1}{2}\vec{\rho}\cdot A\cdot\vec{s}\nonumber\\
 & = & -1+N-\frac{1}{4}\sum_{i,j}\rho_{i}(1-\delta_{ij})s_{j}\nonumber\\
 & = & -1+N-\frac{1}{2}(N-1)\sum_{j}s_{j}\nonumber\\
 & = & 0 \; 
\end{eqnarray}
which is the correct result for a single block. 

Maximum shortening
can also arise in parts of the sequence $\vec{t}$ with some if the
inter-block gaps being $0$. Suppose that $v_{k,k+1}=0$; for any given $k$; then
\begin{eqnarray}
A_{k,k+1} & = & \frac{1}{2}\nonumber\\
A_{i,k} & = & A_{i,k+1}\qquad i\neq k,k+1 \; ,
\end{eqnarray}
and the relevant terms in the expression (\ref{eq:omega_s}) for $\tilde{\omega}$ 
are combined together due to the identity
\begin{equation}
-\frac{1}{4}(s_{k}-2)^{2}-\frac{1}{4}(s_{k+1}-2)^{2}-\frac{A_{k,k+1}}{2}s_{k}s_{
k+1}=-\frac{1}{4}(s_{k}+s_{k+1}-2)^{2}-1 \; ,
\end{equation}
where the $-1$ corresponds to the decreasing the number of the blocks
$N$ to $N-1$. In case of more than one zero gaps this reasoning can be 
applied iteratively.

\subsubsection{The case of small $A$}

Note that
\begin{eqnarray}
A_{k,k-1} & = & A_{k-1,k}=\frac{1-(v_{k}-v_{k-1})}{2}\nonumber\\
A_{kl} & \geq & 0\qquad|k-l|\neq 1 \; ,
\end{eqnarray}
therefore the length $D(\vec{t}\:)$ can be estimated as
\begin{eqnarray}
D(\vec{t}\:) & = & 
v_{N}-v_{1}=\sum_{k=1}^{N-1}v_{k}-v_{k-1}\geq-\sum_{i,j}A_{ij}+N-1\nonumber\\
 & = & -\frac{1}{4}\vec{\rho}A\vec{\rho}+N-1 \; .
\end{eqnarray}
Then our upper estimate of $c$-function exponent becomes 
\begin{equation}
\zeta(\vec{t}\:)\leq2(\tilde{\omega}(\vec{t}\:)-D(\vec{t}\:))\leq-\frac{1}{2}\vec{\rho
}\frac{2A}{1+2A}\vec{\rho}+\frac{1}{2}\vec{\rho}A\vec{\rho}\; ,
\end{equation}
which can be simplified to
\begin{equation}
\zeta(\vec{t}\:)\leq-\frac{1}{2}\vec{\rho}A\frac{1-2A}{1+2A}\vec{\rho}=-\frac{1}{2
}\vec{\rho}A\vec{\rho}+O(A^{2}) \; .
\label{eq:integrality_firstorder}\end{equation}
The leading term is definitely negative, therefore for small deviations
from integrality the $c$-function exponent is negative.

\subsubsection{The case of small shift}

Suppose now that 
\begin{equation}
\vec{s}_{*} =  (1+2A)^{-1}\vec{\rho}
\end{equation}
is still close enough to $\vec{\rho}\;$ so that its rounding is 
identical to the $A=0$ case (\ref{eq:A=00003D0solution}). This always happens if $A$ is small enough; 
however, this is not limited to the approximation used in (\ref{eq:integrality_firstorder}). In this case 
\begin{equation}
\tilde{\omega}(\vec{t}\:)\leq-1+N-\frac{1}{2}\vec{\rho}A(\vec{\rho}+\vec{\delta)}
-\frac{1}{4}\vec{\delta}^{2} \; ,
\end{equation}
and using 
\begin{equation}
D(\vec{t}\:)\geq-\frac{1}{4}\vec{\rho}A\vec{\rho}+N-1 
\end{equation}
one obtains 
\begin{equation}
\zeta(\vec{t}\:)\leq-\frac{1}{4}\vec{\rho}A\vec{\rho}-\frac{1}{2}\vec{\rho}A\vec{
\delta}-\frac{1}{4}\vec{\delta}^{2}  \; .
\end{equation}
Recall that $\rho_{k}=2$ and $A_{ij}>0$ therefore the first
term is negative; the absolute value of the second term is always smaller than or
equal to the first one as $\delta_{k}=\pm 1$. The second term could only cancel
the first one when all $\delta_{k}=-1$, but then the last term is
(maximally) negative. Therefore we again conclude
\begin{equation}
\zeta(\vec{t}\:)<0
\end{equation}
whenever $A\neq 0$.

\subsubsection{Summary of the arguments for integrality}

Our conclusion is that for a cell characterised with a sequence 
\begin{equation}
\vec{t}=(\underbrace{v_{1}\dots v_{1}}_{p_{1}},\underbrace{v_{2}\dots 
v_{2}}_{p_{2}},\dots,\underbrace{v_{N}\dots v_{N}}_{p_{N}})
\end{equation}
to contribute to the roaming $c$-function   
\begin{equation}
v_{k}-v_{k-1}\leq 1
\end{equation}
must be satisfied. Further, any deviation from exact equality to $1$ 
decreases $\zeta$ to first order in the deviation matrix $A$, 
which makes the block's contribution vanish in the roaming limit. Incidentally,
this also explains why the contribution of each cell satisfying the property R 
(\ref{guysrule}) comes from an $O(1)$ region around each centre. For larger deviations
we could prove that they decrease $\zeta$ as long as
the optimal value of  $\vec{s}\;$  is not shifted from $\vec{\rho}$. 

The case of maximum shortening provides an example when decreasing some of the 
$v_{k}-v_{k-1}$ results in a shift of the optimal value of $\vec{s}\;$ such 
that one can again have $\zeta=0$. For a full proof for integrality the only thing left to be 
shown is that these are the only such cases, but for the time being this remains elusive. 

We close with an intuitive picture which is essentially the form factor version 
of the TBA arguments reviewed in Section \ref{sec:stairTBA}. The sinh-Gordon $S$ matrix (\ref{eq:sinhGSmatrix}) 
continued to the roaming trajectory regime (\ref{eq:sinpibper2roaming}) takes the form
\begin{equation}
S(\theta)=\frac{\sinh\theta - i\cosh\vartheta_0}{\sinh\theta + 
i\cosh\vartheta_0} \; .
\end{equation}
It satisfies
\begin{equation}
\mathop{\mathrm{lim}}\limits_{\vartheta_0\rightarrow\infty}
S(\theta\pm\vartheta_0)=\pm\frac{e^\theta - i}{e^\theta + i} \; ,
\end{equation}
and apart from a finite (i.e. $O(1)$) vicinity of the points
$\theta=\pm\vartheta_0$, 
the $S$-matrix is essentially constant:
\begin{equation}
S(\theta)=\begin{cases}
           -1 & |\theta| \ll \vartheta_0 \\
           +1 & |\theta| \gg \vartheta_0 
           \end{cases}
\; .
\end{equation}
The form factor equations 
(\ref{eq:exchangeaxiom},\ref{eq:cyclicaxiom},\ref{eq:kinematicalaxiom}) imply 
that the dependence of the form factor on the rapidity differences is determined 
by the $S$-matrix. Given the above properties of the $S$-matrix it is clear that 
what matters is only whether the $t_{ij}$ (individually) are
smaller, larger or exactly equal to $1$.

The asymptotics of $S$ for $|\theta| \gg \vartheta_0$ are consistent with the 
cluster property (\ref{eq:ffclustering}), which implies that if any of the 
$t_{ij}$ is larger than $1$ (but not necessarily an integer), its exact value 
ceases to have an effect on the asymptotic behaviour of the form factors. The 
asymptotics of $S$ for $|\theta| \ll \vartheta_0$ has a different effect. 
The fact that $S(0)=-1$ means that the form factors vanish whenever any two of their 
rapidity arguments are equal, i.e. they obey a Pauli exclusion principle. One then 
expects that this persists for rapidity differences much smaller than $\vartheta_0$. 
This is indeed true for the minimal form factor function (\ref{eq:minff}), which according 
to the results of Section \ref{sec:roamingminff} behaves as 
\begin{equation}
f(\theta)=\begin{cases}
           0 & |\theta| \ll \vartheta_0 \\
           \rho(\theta-\vartheta_0) & \theta=\vartheta_0 + O(1) \\           
           \bar{\rho}(\theta+\vartheta_0) & \theta=-\vartheta_0 + O(1) \\
           1 & |\theta| \gg \vartheta_0 
           \end{cases}
\end{equation}
for large $\vartheta_0$. 

One can then visualise the situation in the following way. The $c$-function integrand
\begin{equation}
\frac{6+6rE+3r^2E^2+r^3E^3}{2E^4}
|F_{n}^{\Theta}(\theta_1,\dots,\theta_{n})|^2\,e^{-rE}
\end{equation}
can be considered as the negative of a potential function for particles 
positioned at $\theta_1$,$\dots$,$\theta_{n}$. The regions contributing to the 
$c$-function are where the potential is close to being minimal (i.e. the integrand is 
close to being maximal). Due to the above exclusion principle, there is a
short range repulsion between the particles which vanishes when their distance 
is larger than $\vartheta_0$. The $e^{-rE}$ term constrains the particles to 
an interval defined by $|\theta_i|<-\log r$, while the energy denominator tries to 
squeeze them closer to the origin\footnote{As noted in Section 
\ref{sec:cfunction_scaling}, the numerator terms $rE$ do not play any role 
since inside the box forced by the exponential factor they all have an 
upper limit of $O(1)$.}. So the particles would prefer to be placed at 
distances of at least $\vartheta_0$; with the confining potential provided 
by the energy denominator, the minimisation of their potential energy results 
in the formation of ``pockets'' of size $O(1)$ separated by distances 
$\vartheta_0$, which is exactly the content of the integrality assumption. Once
this is realised, one can proceed to find the optimal distribution of particles 
between the pockets following the method outlined in Sections \ref{sec:naive_FF_powercounting} 
and \ref{sec:cfunction_scaling}, which results in the rule (\ref{guysrule}).

\resection{Effective magnonic system}

Here we show that form factor relocalisation permits the
reconstruction of form factors in the massless flows between the
minimal models induced by the $\Phi_{1,3}$ perturbation. The basic
idea is very simple. Using Theorem R formulated in section
\ref{cellsandloc} the c-theorem sum
rule (\ref{c-sumrule}) can be rewritten as a sum over the contributing
cells $[a_1,\dots,a_n]$
\begin{eqnarray}
c(r)
&=& 3
\sum_{n=0}^{\infty}\sum\limits_{[a_1,\dots,a_n]}\;
\int\limits_{[a_1,\dots,a_n]}\!\frac{d\theta_1\dots d\theta_{n}}
{(2\pi)^n(n!)}\,
\frac{6+6rE+3r^2E^2+r^3E^3}{2E^4}\,
|m^{-2}F_{n}^{\Theta}(\theta_1\dots\theta_{n})|^2\,e^{-rE}\nonumber\\
&&+\dots \; ,
\end{eqnarray}
where the omitted terms decay exponentially with increasing $\vartheta_0$.

Each contributing cell can be reparameterised by integration variables
$\beta_i$ that have the centre of the cell as their origin:
\begin{equation}
 \theta_i=\beta_i+a_i\frac{\vartheta_0}{2}\qquad i=1,\dots,n\;. 
\end{equation}
Using the rescaling 
\begin{equation}
 r=\tilde{r}e^{-(k-1)\vartheta_0/2}
\end{equation}
to localise the integral to step $k$ one obtains the $c$-function for the flow corresponding to 
step $k$ as
\begin{eqnarray}
c_k(r)&=&
\lim_{\vartheta_0\rightarrow\infty}
\sum_{n=0}^{\infty}
\sum\limits_{\genfrac{}{}{0pt}{}{[ a_1,\dots,a_n ]}{a_n = k-1}}\;
\int\limits_{[a_1,\dots,a_n]}\!\frac{d\theta_1\dots d\theta_{n}}
{(2\pi)^n(n!)}\,
\Bigg[\frac{6+6rE+3r^2E^2+r^3E^3}{2E^4}\,e^{-rE}\nonumber\\
&& \times |m^{-2}F_{n}^{\Theta}(\beta_1+a_1\vartheta_0/2,\dots,\beta_n+a_n\vartheta_0/2)|^2\Bigg] 
\nonumber\\
&&E=\sum\limits_{i=1}^{n}\cosh(\beta_i+a_i\vartheta_0/2)
\; .
\end{eqnarray}
where the sum runs over sequences $[a_1,\dots,a_n]$ allowed by the rule expressed in Theorem R. 
Rewriting the allowed sequences to block notation and 
relabelling the $\beta_i$ accordingly as follows
\begin{eqnarray}
 [a_1,\dots,a_n]=[\underbrace{-k+1,\dots,-k+1}_{2 p_{-k+1}},
 \underbrace{-k+3,\dots,-k+3}_{2 p_{-k+3}},\dots,\underbrace{k-1,\dots,k-1}_{2 p_{k-1}}]
\nonumber\\
 \{\beta_1,\dots\,\beta_n\}=\{\beta^{(-k+1)}_1,\dots,\beta^{(-k+1)}_{2 p_{-k+1}},
 \beta^{(-k+3)}_1,\dots,\beta^{(-k+3)}_{2 p_{-k+3}},\dots,
 \beta^{(k-1)}_1,\dots,\beta^{(k-1)}_{2 p_{k-1}}
 \}\; ,
\end{eqnarray}
the $c$-function for step $k$ takes the form
\begin{eqnarray}
c_k(r) &=&
\sum_{n=0}^{\infty}
\sum\limits_{\genfrac{}{}{0pt}{}{\{p_{i}\}}{2\sum\limits_i p_i=n}}\;
\frac{1}{n!}
\int\limits\!\left(\prod\limits_{i=-k+1}^{k-1} \prod\limits_{j_i=1}^{2p_i}\frac{d\beta^{(p_i)}_{j_i}}{2\pi}\right)\,
\Bigg[ \frac{6+6\tilde{r}\tilde{E}+3\tilde{r}^2\tilde{E}^2+\tilde{r}^3\tilde{E}^3}{2\tilde{E}^4}
\,e^{-\tilde{r}\tilde{E}}
\nonumber\\
&& \times \left|\mathcal{F}_{\{p_i\}}^{\Theta}\left(
\beta^{(-k+1)}_1,\dots,\beta^{(-k+1)}_{2 p_{-k+1}},
 \beta^{(-k+3)}_1,\dots,\beta^{(-k+3)}_{2 p_{-k+3}},\dots,
 \beta^{(k-1)}_1,\dots,\beta^{(k-1)}_{2 p_{k-1}}
\right)\right|^2\Bigg] 
\nonumber\\
&&\tilde{E}=\sum\limits_{i=1}^{2p_{-k+1}}e^{-\beta^{(p_{-k+1})}_i}
+\sum\limits_{i=1}^{2p_{k-1}}e^{\beta^{(p_{k-1})}_i}
\; ,
\label{magnonic-c-sumrule}
\end{eqnarray}
where $\{p_i\}$ is a shorthand for $\{p_{-k+1},p_{-k+3},\dots,p_{k-1}\}$ and 
\begin{eqnarray}
&&\mathcal{F}_{\{p_i\}}^{\Theta}\left(
\beta^{(-k+1)}_1,\dots,\beta^{(-k+1)}_{2 p_{-k+1}},
 \beta^{(-k+3)}_1,\dots,\beta^{(-k+3)}_{2 p_{-k+3}},\dots,
 \beta^{(k-1)}_1,\dots,\beta^{(k-1)}_{2 p_{k-1}}
\right)=\nonumber\\
&&\lim_{\vartheta_0\rightarrow\infty}
m^{-2}e^{-2(k-1)\vartheta_0}
F_{n}^{\Theta}(\beta_1+a_1\vartheta_0/2,\dots,\beta_n+a_n\vartheta_0/2)
\; ,
\label{roamingFF}
\end{eqnarray}
where on the right hand side we returned the original notation for the cell coordinates and the shifted 
rapidity variables $\beta_i$ in order to keep the formula compact.

The result (\ref{magnonic-c-sumrule}) has a very compelling interpretation in terms of the massless flow 
that corresponds to step $k$. Namely, the rapidities $\beta^{(l)}_i$ can be thought of as being grouped into 
$k$ bins corresponding 
to their upper index. The ones in the leftmost bin correspond to left-moving particles, 
while the ones in the rightmost bin to right-moving particles, both massless. This is reflected by their 
contribution to the rescaled dimensionless energy $\tilde{E}$. Note, however, that the rapidities in
intermediate bins do not contribute to the energy at all. Recalling that the scattering theory 
of the massless flow is described by a factorised scattering theory of 
massless kinks (corresponding to domain walls), one can understand the rapidity variables in the $k-2$ 
intermediate bins as magnonic degrees of freedom describing the multiplet structure of multi-kink states. 
Note that the number and the 
adjacency structure of these bins exactly conforms to the Dynkin diagram encoding the TBA system derived by 
Zamolodchikov for the corresponding massless flow \cite{Zamolodchikov:1991vh}.

Therefore the functions $\mathcal{F}^\Theta$ can be interpreted as form factors of the trace of the 
stress energy tensor (expressed in units $m=1$) along the massless flow in the magnonic basis of multi-kink states. 
In principle such form factors could be obtained from the form factor bootstrap; however, the non-diagonal nature of 
kink scattering makes it notoriously difficult to obtain solutions of the bootstrap equations. Indeed, the only 
nontrivial flow for 
which the form factor solution is presently known is the case $k=2$ corresponding to the flow from the 
tricritical to critical Ising model, where the scattering is effectively diagonal and therefore magnons 
are absent. By a somewhat tedious, but elementary computation it can be shown that the form factors constructed 
according to (\ref{roamingFF}) for $k=2$ indeed match the ones obtained by Delfino et al. \cite{Delfino:1994ea}, 
up to a phase factor which is irrelevant for the spectral weight.

\resection{Conclusions}

In the present work we have shown how Zamolodchikov's roaming flows can be 
analysed via the $c$-theorem, representing the $c$-function as 
a form factor spectral sum. In particular, we demonstrated that the well-known 
staircase structure of the free energy obtained from the TBA can be recovered 
from the $c$-theorem. 

We have shown that this is the result of an interesting property of sinh-Gordon 
form factors under the roaming continuation, which results in a 
relocalisation of multi-particle contributions 
to the $c$-function. Our demonstration is in fact an almost complete 
mathematical 
proof, the only ingredient not fully proven is the integrality assumption 
(\ref{integrality}); however, we presented quite strong evidence that this  
property holds as well.

The relocalisation property, central to our results, is essentially
equivalent to the construction of exact form factors for the trace of
energy-momentum tensor in the field theories describing interpolating
flows between minimal models. This is a very interesting result given
that solving the form factor bootstrap for these flows is a rather
nontrivial task due to the non-diagonal nature of the exact
$S$-matrix. The scattering matrix in these theories is non-diagonal,
which prevents finding the form factors via a general Ansatz involving
polynomial recursion relations
as in the case of diagonal models. In some field theories with
non-diagonal scattering, the form factor bootstrap can still be solved
using some ingenious methods (cf. e.g.
\cite{Smirnov:1992vz,Lukyanov:1993pn,Babujian:1998uw}); 
however, no generally applicable approach exists. Constructing form
factors using relocalisation of simple form factor solutions under a
roaming continuation effectively circumvents this problem, and the
resulting form factors can be used to compute correlation functions.
For the trace of energy-momentum operator this is equivalent to
constructing Zamolodchikov's $c$-function for the flow; however, there
is much more potential in the approach, as we discuss below.

One possible extension of our results would be to examine the
properties of other form factor solutions, corresponding to
exponential operators $e^{\kappa b\Phi}$ with exponents $\kappa\neq
\pm 1$.  Taking the roaming limit of such form factors, one expects to
find the form factors of other relevant local fields of $\Phi_{1,3}$
perturbed minimal conformal field theories, thus giving access to
further interesting correlators such as e.g. magnetic susceptibility.
The $\Delta$-theorem \cite{Delfino:1996nf} evaluated as a sum rule
over form factors could help in operator identification and also give
a more detailed characterisation of the renormalisation group flow by
relating the operator contents at the endpoints of the flow.

It is very plausible that the relocalisation property extends to other
interesting theories such as homogeneous sine-Gordon models where
staircase structures corresponding to resonance spectra have been
found 
\cite{Miramontes:1999hx,CastroAlvaredo:1999em,Dorey:2004qc,
CastroAlvaredo:2000ag,CastroAlvaredo:2000nr},
and to already known extensions of
roaming trajectories 
\cite{Martins:1992ht,Dorey:1992bq,Martins:1992sx,Dorey:1992pj}.  
In this connection the
association of the form factor structures to Dynkin-like structures is
also interesting to investigate and clarify, as they can lead to a
classification of form factor solutions for a wide class of scattering
theories. Such Dynkin form factors would allow for a direct way to
conjecture form factor solutions in the case of non-diagonal
scattering. This would be similar in spirit to the Dynkin TBAs of
\cite{Ravanini:1992fi}; in fact, based on the present work one expects
a deep relation to the structure of TBA diagrams. In particular, note
that the form factors constructed via relocalisation account for the
non-diagonal nature of scattering theory via magnonic particles, which
is exactly analogous to the TBA case.

\bigskip

\subsection*{Acknowledgements}
We are grateful to 
Olalla Castro-Alvaredo,
Benjamin Doyon, 
Luis Miramontes
and Fedor Smirnov for useful discussions, and 
especially to Yaroslav Pugai for suggesting we utilise the 
Lukyanov representation for the form factors. 
P.E.D. acknowledges support from the STFC under the Consolidated Grants
ST/J000426/1 and ST/L000407/1, G.S. was supported by an STFC
studentship, and G.T. was supported by 
Lend\"ulet grant LP2012-50/2014.
This research was also supported in part by the Marie Curie network
GATIS (gatis.desy.eu) of the European Union's Seventh Framework
Programme FP7/2007-2013/ under REA Grant Agreement No 317089.

\appendix

\resection{Exact form factors in sinh-Gordon theory}\label{sinhGFF}
\label{app:shGFF}

The form factors of the exponential operators
\begin{equation}
F_{n}^{(\kappa)}(\theta_{1},\theta_{2},\dots,\theta_{n})=\langle0|e^{\kappa b\Phi}|\theta_{
1},\dots,\theta_{n}\rangle\label{eq:expffdef}
\end{equation}
in the sinh-Gordon model (\ref{eq:sinhGaction})
are particular solutions of the following form factor bootstrap equations:\\

I. Lorentz invariance:
\begin{equation}
 F_{n}^{\mathcal{O}}(\theta_{1}+\Lambda,\dots,\theta_{n}+\Lambda)=
 e^{s_{\mathcal{O}}\Lambda} F_{n}^{\mathcal{O}}(\theta_{1},\dots,\theta_{n})
 \, ;
 \label{eq:Lorentzaxiom}
\end{equation}

II. Exchange:
\begin{equation}
 F_{n}^{\mathcal{O}}(\theta_{1},\dots,\theta_{k},\theta_{k+1},\dots,\theta_{n})=
 S(\theta_{k}-\theta_{k+1})F_{n}^{\mathcal{O}}(\theta_{1},\dots,\theta_{k+1},\theta_{k},\dots,\theta_{n})
 \, ;
 \label{eq:exchangeaxiom}
\end{equation}

III. Cyclic permutation: 
\begin{equation}
F_{n}^{\mathcal{O}}(\theta_{1}+2i\pi,\theta_{2},\dots,\theta_{n})=
F_{n}^{\mathcal{O}}(\theta_{2},\dots,\theta_{n},\theta_{1})
\, ;
\label{eq:cyclicaxiom}
\end{equation}

IV. Kinematical singularity
\begin{equation}
-i\mathop{\textrm{Res}}_{\theta=\theta^{'}}
F_{n+2}^{\mathcal{O}}(\theta+i\pi,\theta^{'},\theta_{1},\dots,\theta_{n})=
\left(1-\delta_{i\, j}\prod_{k=1}^{n}S_{i\, i_{k}}(\theta-\theta_{k})\right)
F_{n}^{\mathcal{O}}(\theta_{1},\dots,\theta_{n})
\, ;
\label{eq:kinematicalaxiom}
\end{equation}
with Lorentz spin $s_{\mathcal{O}}=0$. The general solution of these equations has the form \cite{Fring:1992pt}
\begin{eqnarray}
F_{n}^{(\kappa)}(\theta_{1},\theta_{2},\dots,\theta_{n}) & = &
\mathcal{G}_{\kappa}\left(\frac{4\sin\gamma
}{\mathcal{N}}\right)^{n/2}\frac{Q_{n}^{(\kappa)}(x_{1},\dots,x_{n})}{
\prod\limits
_{r<s}(x_{r}+x_{s})}\prod_{r<s}f(\theta_{r}-\theta_{s}) \; ,\nonumber \\
 &  & x_{r}=e^{\theta_{r}} \; ,
 \label{eq:KMff}
\end{eqnarray}
where the minimal two-particle form factor is
\eq
f(\theta)  = 
\mathcal{N}\exp\left[8\int_{0}^{\infty}\frac{dt}{t}\sin^{2}
\left(\frac{ t(i\pi-\theta)}{2\pi}\right)
\frac{\sinh\frac{\gamma}{2\pi} 
\sinh(1-\frac{\gamma}{\pi})\frac{t}{2}\sinh\frac{t}{2}}{\sinh^{2}t}\right]\,.
\label{eq:minff}
\en
The normalisation constant $\mathcal{N}$, chosen so that
$ f(\pm\infty)=1$, is
\eq
\mathcal{N}=\exp\left[-4\int_{0}^{\infty}\frac{dt}{t}
\frac{\sinh\frac{\gamma}{2\pi}
\sinh(1-\frac{\gamma}{\pi})\frac{t}{2}\sinh\frac{t}{2}}{\sinh^{2}t}\right]
\en
and
\begin{equation}
\mathcal{G}_{\kappa}=\langle0|e^{\kappa b\Phi}|0\rangle\label{eq:expvev}
\end{equation}
is the vacuum expectation value of the field. An exact formula for the
expectation values of exponential fields in the sine-Gordon model
was obtained in \cite{Lukyanov:1996jj}, from which the relevant
expectation values in the sinh-Gordon theory can be obtained by analytic
continuation; however, their explicit forms are not needed in the sequel. 
The symmetric polynomials
$Q_{n}$ satisfy the recursive relations \cite{Fring:1992pt}
\begin{equation}
(-1)^{n}Q_{n+2}(-x,x,x_{1},\dots,x_{n})  = 
xD_{n}(x|x_{1},\dots,x_{n})Q_{n}(x_{1},\dots,x_{n})\; ,
\label{eq:Qrecursion}
\end{equation}
where
\begin{align}
&D_{n}(x|x_{1},\dots,x_{n})  =  \nonumber\\
&\qquad\frac{1}{2\sin\gamma}\sum_{l,k=0}^{n}(-1)^{l}\sin\left((k-l)\gamma   
\right)x^{2n-l-k}\sigma_{l}^{(n)}(x_{1},\dots,x_{n})\sigma_{k}^{(n)}(x_{1},
\dots,x_{n})
\; ,
\end{align}
and $\sigma_{l}^{(n)}$ denotes the elementary symmetric polynomials 
defined by
\begin{eqnarray}
\prod_{r=1}^{n}(x+x_{r}) & = &
\sum_{l=1}^{n}x^{n-l}\sigma_{l}^{(n)}(x_{1},\dots,x_{n})\nonumber \\
\sigma_{l}^{(n)}\equiv0 &  & \mbox{if }\: l<0\mbox{ or
}l>n\,.
\label{eq:sympolydef}
\end{eqnarray}
For the exponential operators (\ref{eq:expffdef}), the polynomials $Q_n$ are given explicitly 
by the following determinant representation
\cite{Koubek:1993ke}
\begin{eqnarray}
Q_{1}^{(\kappa )} & = & [\kappa ]\nonumber \\
Q_{n}^{(\kappa )} & = & [\kappa ]\det M^{(n)}(k)\qquad n>1\nonumber \\
 &  &
M_{rs}^{(n)}(\kappa )=[r-s+\kappa ]\sigma_{2r-s}^{(n)}(x_{1},\dots,x_{n})\quad
r,s=1,\dots,n-1\nonumber \\
 &  & [a]=\frac{\sin\,a\gamma}{\sin\gamma}\,.
\label{eq:kmdet}
\end{eqnarray}
Another useful representation of the same form factor can be obtained
from Lukyanov's result for breather form factors in the sine-Gordon
model \cite{Lukyanov:1997bp}. Analytic continuation of first breather
form factors to imaginary sine-Gordon coupling gives the following
result for form factors of exponential operators in sinh-Gordon model:
\begin{align}
F_{n}^{(\kappa )}\left(\theta_{1},\dots,\theta_{n}\right) & = \,
\mathcal{G}_{\kappa }\frac{i^{n}}{\left(\mathcal{N}\sin\gamma
\right)^{n/2}}P_{n}^{(\kappa )}\left(\theta_{1},\dots,\theta_{n}\right)
\prod_{r<s}f(\theta_{r}{-}\theta_{s})\,,
\label{eq:lukyanov_rep}\\[3pt]
P_{n}^{(\kappa )}\left(\theta_{1},\dots,\theta_{n}\right) & = 
\sum_{\{\alpha_{j}=\pm1\}}\left\{
\left(\prod_{j=1}^{n}\alpha_{j}e^{-i\alpha_{j}
\kappa \gamma}\right)\prod_{r<s}
\left(1-i\frac{\alpha_{r}{-}\alpha_{s}}{2}\frac{\sin\gamma
}{\sinh\left(\theta_{r}{-}\theta_{s}\right)}\right)\right\}.
\nonumber
\end{align}
From this representation it is easy to see that the form factors of 
exponential operators satisfy the cluster property \cite{Delfino:1996nf}
\begin{eqnarray}
&&F_{n+l}^{(\kappa )}\left(\theta_{1}+\Lambda,\dots,\theta_{n}+\Lambda,
\theta_{1}',\dots,\theta_{l}'\right)
= \frac{1}{\mathcal{G}_{k}}
F_{n}^{(\kappa )}\left(\theta_{1},\dots,\theta_{n}\right)
F_{l}^{(\kappa )}\left(\theta_{1}',\dots,\theta_{l}'\right)
+O\left(e^{-\Lambda}\right) \nonumber\\
&&\; \mathrm{when }\;\Lambda\rightarrow\infty .
\label{eq:ffclustering}
\end{eqnarray}
The identification between the two representations is given by the
relation
\begin{equation}
P_{n}^{(\kappa )}\left(\theta_{1},\dots,\theta_{n}\right)=\left(-2i\sin\gamma
\right)^{\!n\,}\frac{Q_{n}^{(\kappa )}(e^{\theta_{1}},\dots,e^{\theta_{n}})}%
{\prod\limits_{r<s}\left(e^{\theta_{r}}+e^{\theta_{s}}\right)}\,.
\label{eq:km_lukyrel}
\end{equation}
For the $c$-function we need the form factors of the trace of the
energy momentum tensor $\Theta$, which is given by \cite{Fring:1992pt}
\begin{equation}
F_{n}^{\Theta}\left(\theta_{1},\dots,\theta_{n}\right)=\begin{cases}
\frac{2\pi
m^{2}}{F_{2}^{(1)}(i\text{\ensuremath{\pi},0})}F_{n}^{(1)}\left(\theta_{1},\dots
,\theta_{n}\right)
& \quad n\mbox{ even}\\
0 & \quad n\mbox{ odd}
\end{cases}\label{eq:Theta0exp1}
\end{equation}
and the cluster property (\ref{eq:ffclustering}) can be rewritten as
\begin{eqnarray}
F_{n+l}^{\Theta}\left(\theta_{1}+\Lambda,\dots,\theta_{n}+\Lambda,
\theta_{1}',\dots,\theta_{l}'\right)
&=& \frac{2\sin\gamma}{\pi m^2}\,
F_{n}^{\Theta}\left(\theta_{1},\dots,\theta_{n}\right)
F_{l}^{\Theta}\left(\theta_{1}',\dots,\theta_{l}'\right)
\nonumber\\
&&+O\left(e^{-\Lambda}\right)\;, 
\label{eq:Thetaclustering}
\end{eqnarray}
provided both $n$ and $l$ are even.

%
%

\end{document}

%% file: Stairflow.pdf_t
\begin{picture}(0,0)%
\includegraphics{Stairflow.pdf}%
\end{picture}%
\setlength{\unitlength}{3947sp}%
\begingroup\makeatletter\ifx\SetFigFont\undefined%
\gdef\SetFigFont#1#2#3#4#5{%
  \reset@font\fontsize{#1}{#2pt}%
  \fontfamily{#3}\fontseries{#4}\fontshape{#5}%
  \selectfont}%
\fi\endgroup%
\begin{picture}(8333,3198)(2686,-5053)
\put(6901,-2086){\makebox(0,0)[lb]{\smash{{\SetFigFont{17}{20.4}{\sfdefault}{\mddefault}{\updefault}{\color[rgb]{0,0,0}Increasing $\Theta_0$}%
}}}}
\end{picture}%